\documentclass{article}

\usepackage{arxiv}

\usepackage[utf8]{inputenc} % allow utf-8 input
\usepackage[T1]{fontenc}    % use 8-bit T1 fonts

\usepackage{enumerate,latexsym}
\usepackage{amsmath,amssymb,amsthm,amsfonts,graphicx}
\usepackage{color,epic} %natbib}
\usepackage{url}            % simple URL typesetting
\usepackage{hyperref} 
\usepackage{fullpage}
\usepackage{array}
\usepackage{arydshln}
\usepackage{cleveref}
\usepackage{algorithm}
\usepackage{algpseudocode}
\usepackage{threeparttable}
\usepackage{tikz}
\usepackage{pgfplots}
\usepgfplotslibrary{fillbetween}
\pgfplotsset{compat=1.16}
\pgfplotsset
{
    legend image with text/.style={
        legend image code/.code={%
            \node[anchor=center] at (0.3cm,0cm) {#1};
        }
    },
}

\definecolor{burntorange}{RGB}{191, 87, 0}
\definecolor{burntblue100}{RGB}{0, 95, 134}
\definecolor{burntgreen100}{RGB}{67, 105, 91}
\definecolor{graphred}{RGB}{160, 82, 71}
\definecolor{graphgreen}{RGB}{112, 173, 71}
\definecolor{graphblue}{RGB}{47, 82, 143}
\definecolor{basegreen}{RGB}{65, 117, 25}
\definecolor{basecyan}{RGB}{105, 183, 184}
\definecolor{baseyellow}{RGB}{182, 183, 41}
\definecolor{tabolive}{RGB}{188, 189, 134}
\definecolor{taborange}{RGB}{255, 127, 14}
\definecolor{tabpink}{RGB}{227, 119, 194}
\definecolor{lightsteelblue}{RGB}{176, 189, 217}
\definecolor{steelblue}{RGB}{70, 130, 180}
\definecolor{deepskyblue}{RGB}{0, 191, 255}
\definecolor{darkgreen}{RGB}{50, 90, 19}
\definecolor{seagreen}{RGB}{46, 139, 87}
\definecolor{cssgreen}{RGB}{0, 128, 0}
\definecolor{dimgray}{RGB}{96, 96, 96}
\definecolor{chocolate}{RGB}{210, 105, 30}
\definecolor{csscyan}{RGB}{0, 255, 255}
\definecolor{palevioletred}{RGB}{186, 106, 137}
\definecolor{tan}{RGB}{194, 172, 133}
\definecolor{slateblue}{RGB}{95, 84, 195}
\definecolor{cssorange}{RGB}{227, 158, 37}
\definecolor{limegreen}{RGB}{116, 197, 61}
\definecolor{crimson}{RGB}{178, 35, 55}
\definecolor{royalblue}{RGB}{65, 105, 225}
\definecolor{lightskyblue}{RGB}{135, 206, 250}
\definecolor{indianred}{RGB}{205, 92, 92}
\definecolor{lightcoral}{RGB}{240, 128, 128}
\definecolor{darkgoldenrod}{RGB}{184, 134, 11}
\definecolor{goldenrod}{RGB}{255, 215, 0}
\definecolor{mediumblue}{RGB}{0, 0, 205}
\definecolor{darkviolet}{RGB}{148, 0, 211}
\definecolor{csslime}{RGB}{0, 255, 0}
\definecolor{darkcyan}{RGB}{0, 139, 139}
\definecolor{cssolive}{RGB}{128, 128, 0}
\definecolor{cornflowerblue}{RGB}{100, 149, 237}
\definecolor{csspurple}{RGB}{128, 0, 128}

\providecommand{\keyword}[1]{\textbf{\textit{Keywords:}} #1}

\def\R{{\rm\vrule depth0ex width.4pt\kern-.08em R}}
\def\reals{{\rm\vrule depth0ex width.4pt\kern-.08em R}}
\def\Prob{{\rm\vrule depth0ex width.4pt\kern-.08em P}}

\newcommand{\theauthors}{Yanyue Ding, Sudesh K.\ Agrawal, Jincheng Cao, Lauren Meyers, John J.\ Hasenbein}

\newcommand{\thetitle}{Surveillance Testing for Rapid Detection of Outbreaks in Facilities}
\newcommand{\therunningtitle}{\thetitle}

\newcommand{\thekeywords}{COVID-19, surveillance testing, agent-based model, contact networks, nursing homes}

\title{\thetitle}

\date{} 					% Or removing it

\author{ Yanyue Ding \\
    Graduate Program in Operations Research \& Industrial Engineering (ORIE) \\
    University of Texas at Austin\\
	Austin, Texas 78712 \\
	\AND
    \href{https://sudeshkagrawal.github.io}{Sudesh K.\ Agrawal} \\
    Graduate Program in ORIE \\
	University of Texas at Austin\\
	Austin, Texas 78712 \\
	\And
    Jincheng Cao \\
    Department of Mathematics \\
	University of Texas at Austin\\
	Austin, Texas 78712 \\
	\AND
    Lauren Meyers \\
    Graduate Program in ORIE, and \\ 
    Department of Integrative Biology \\
	University of Texas at Austin\\
	Austin, Texas 78712 \\
	\And
	\href{http://sites.utexas.edu/hasenbein/}{John J.\ Hasenbein} \\
	Graduate Program in ORIE \\
	University of Texas at Austin\\
	Austin, Texas 78712 \\
	\texttt{\href{mailto:jhas@mail.utexas.edu}{jhas@mail.utexas.edu}} \\
}

\renewcommand{\shorttitle}{\therunningtitle}

\usepackage{hyperref}       % hyperlinks
\hypersetup
{
    pdfauthor={\theauthors},
    pdftitle={\thetitle},
    pdfsubject={A preprint for arXiv},
    pdfkeywords={\thekeywords},
    pdfborderstyle={/S/S/W 0},
    linkbordercolor=white,
    citebordercolor=white,
    urlbordercolor=white,
    colorlinks=false,
}

\begin{document}

\maketitle

\begin{abstract}
\noindent This paper develops an agent-based disease spread model on a contact network in an effort to guide efforts at surveillance testing in small to moderate facilities such as nursing homes and meat-packing plants. The model employs Monte Carlo simulations of viral spread sample paths in the contact network. The original motivation was to detect COVID-19 outbreaks quickly in such facilities, but the model can be applied to any communicable disease. In particular, the model provides guidance on how many test to administer each day and on the importance of the testing order among staff or workers. 
\end{abstract}

%\begin{keyword}[class=AMS]
%\kwd[Primary ]{60K35}
%\kwd{60K35}
%\kwd[; secondary ]{60K35}
%\end{keyword}

\keyword{COVID-19, surveillance testing, agent-based model, contact networks, nursing homes}

%\begin{keyword}
%\kwd{Naor's model, parameter uncertainty, unobservable queues, social optimization,
%revenue maximization}
%\kwd{\LaTeXe}
%\end{keyword}

\section{Introduction}
In this paper, we develop an agent-based disease spread model on a contact network to optimize surveillance testing in
facilities. Generally speaking, the model's scope involves facilities of small to moderate size, say, with dozens to hundreds of employees. 
The original motivation derives from the need to quickly detect outbreaks of COVID-19 in facilities such as nursing homes, factories,
and meat-packing plants, in which close contact is nearly unavoidable. Our focus is on \emph{surveillance testing} or proactive testing
of staff. In other words, the premise is that there is not currently an outbreak in the facility, and the manager wishes to proactively 
test asymptomatic staff in order to quickly detect, and prevent, an outbreak.  Given a contact network,
along with other modeling parameters detailed in the sequel, the primary goal is to answer the following:

\begin{enumerate}
\item How many employees should be tested each day in order to detect an outbreak with a given probability?
\item What is the optimal sequence in which to test employees? 
\end{enumerate}

We propose a relatively simple Monte Carlo simulation model that is tailored to answering the two questions. Our model is
similar to other agent-based SEIR models in the literature.  However, the novelty is in our focus on optimization and decision making,
and how it relates to the structure of the contact network. In the first part of the paper, we examine generic contact networks that
may be of use when there is little data on the interactions in a facility. In the second part of the paper, we present more nuanced contact
models, that are derived from the staff and interaction structure in nursing homes. Such models are preferred when there is detailed
data on a facility's operations. 

\section{Literature Review}

COVID-19 is a critical health issue worldwide and is likely to continue to be of concern for many years. 
%In 2020, COVID-19 became a critical public health issue world-wide and is likely to continue to be of concern for several years. 
Even with vaccine development, it is imperative to detect outbreaks at an early stage to develop effective mitigation measures. According to some studies, 38\%--80\% of cases are pre-symptomatic or asymptomatic in Long-Term-Care Centers (LTCs) \cite{Bernadou2021,Yen2020}. Moreover, 56\%--95.5\% of disease transmissions are caused by  asymptomatic cases \cite{Ferretti2020,Harada2020}. Early in the COVID-19 pandemic, in many facilities, only individuals with COVID-like symptoms were tested, due to limited testing resources. However, several studies show that screening symptomatic cases only is not enough to control outbreaks \cite{danis2020high,Garibaldi2021,Harada2020}. Starting in March 2020, the Centers for Disease Control and Prevention (CDC) and the Centers for Medicare \& Medicaid Services (CMS) offered various COVID-19 surveillance testing strategies to hospitals, nursing homes, schools, and other facilities, as the pandemic progressed. Now, surveillance testing is a widely applied strategy to contain the spread of the virus, especially for asymptomatic cases \cite{Bai2017,Drenkard2021}. Apart from suggestions provided by the government, managers and researchers in various facilities developed their own surveillance testing strategies, combined with contact interventions, based on the facility's contact patterns and healthcare resources. In LTCs for example, the strategies include daily individual health screening via surveys or front desk checks \cite{Baek2020,Garibaldi2021}, daily temperature and oxygen level checks \cite{danis2020high}, serological anti-body checks \cite{Strand2021}, and reverse transcription-polymerase chain reaction (RT-PCR) tests.

In terms of evaluation of testing strategies, some strategies are evaluated by outbreak case studies or analysis of publicly recorded data \cite{Albogami2021,ChenHJ2021,Leal-Neto2020,McMichael2020,Sudre2021}. In addition, strategies may be evaluated using variations of the stochastic susceptible-exposed-infected-recovered/removed (SEIR) model. 
Some researchers have made modifications to the traditional SEIR model, to account for nuances of COVID-19.
%Compared with the traditional differential equation-based SEIR model, recent modifications have been made for COVID-19. 
These include adding, removing, or subdividing the standard stages in the network. For example, the exposed stage $E$ is sometimes removed, and the remaining SIR model is used \cite{Ciaperoni2020,Gaeta2021}. Saha \cite{Saha2021} sub-divides the infected stage $I$ into detected and undetected infected stages. Chapman et al.\ \cite{Chapman2021} assume all infected individuals experience a pre-symptomatic stage, after which the corresponding node in the model transitions to an asymptomatic, mild symptomatic or severe symptomatic stage. Another innovation is to allow the parameters in the differential equations defining the SEIR model to be non-linear. Chen and Wei \cite{Chen2020} and Shu et al.\ \cite{Shu2012} assume that the parameters in the differential equations can be non-linear functions of time. 
Some studies modify parameters and the contact network structure over time. These adaptations are primarily used in comparing and finding the best surveillance screening and intervention strategies. Hou et al.\ \cite{Hou2020} and Shaikh et al.\ \cite{Shaikh2020} estimate the parameter settings under various surveillance testing and intervention efficacy assumptions and perform a parametric analysis. Ciaperon et al.\ \cite{Ciaperoni2020} study the dynamic SEIR model by temporarily removing the edges in the contact network due to quarantine intervention for individuals who have COVID-like symptoms or tested positive. 

One key assumption in the differential-equation-based SEIR models is that the population is well-mixed and features are homogeneous among individuals \cite{Chapman2021,Chen2020,Chowell2016}. The standard SEIR model might be suitable for predicting outbreak progression in a large well-mixed population while it might over-simplify dynamics in special networks \cite{Chowell2016}. Broadly speaking, two ways to address these issues have been developed in the literature. One is to classify individuals who share similar characteristics into subgroups and use specific differential parameters in each subgroup. For example, Besse and Faye \cite{Besse2021} divide the population by spatial features and use a standard SIR model with heterogeneous parameters in each spatial group. The disease transmission between groups is estimated by heat equations. Shaikh et al.\ \cite{Shaikh2020} divide the population in an LTC  by the occupations, and use different transmission probabilities for each group. A second solution, requiring significant computational effort, is to use  network-based stochastic models that model each individual. 
%Watt and Strogatz \cite{Watts1998} modify  symmetric networks by randomly re-connecting edges between nodes. They call the modified network a “small-world” and found that the disease can spread more effectively such networks. 
Ames et al.\ \cite{Ames2011} believe human social networks tend to be clustered and heterogeneous. In particular, they study disease transmission features in networks with various coefficient and clustering properties. Their results show that the structure of the network has a profound influence on disease progression. Similarly, Meyers et al.\ \cite{Meyers2005} test intervention strategies for severe acute respiratory syndrome (SARS) in various networks. Their analysis indicates that contact patterns in the network impact the size of epidemic significantly, even if the basic reproduction rates are the same among different networks. Herrera et al.\ \cite{Herrera2016} study surveillance strategies in different networks and find that the best surveillance strategies are based on network structures.

 Rennert et al.\ \cite{Rennert2021} modify the SIR model by distinguishing symptomatic and pre-symptomatic stages and adding an isolation stage. Their results show that voluntary and random testing is not enough to protect students and faculty on a campus. They recommend performing rigorous testing during the semester. They also provide a novel solution by continuing to test the target group if a positive case in random surveillance occurs, which can be more effective in controlling disease spread. Chapman et al.\ \cite{Chapman2021} applied the SEIR model to simulate COVID-19 transmission in a congregated shelter population. The model was tested in shelters in San Francisco, Boston, and Seattle. The results showed that a combined strategy that involves daily screening, twice-weekly PCR testing and contact intervention is the best strategy to detect and avert an outbreak at an early stage. Smith et al.\ \cite{Smith2020best,Smith2020} used an individual-based SEIR model to simulate the transmission of COVID-19 based on detailed contact data in an LTC in order to find the best surveillance strategy. They compared the strategies when daily testing capacities of 1, 2, 4, 8, 16, and 32 are available. 
 
Apart from COVID-19-based models, other researchers have discussed the importance of considering the structure of contact networks when designing surveillance strategies. For example, Herrera-Diestra et al.\ \cite{herrera2021network} study contact networks modeling the spread of foot-and-mouth
disease among cattle farms in Turkey, and note that such networks have typical ``small-world'' properties. Although they suggest some general surveillance principles based on their observations, they do not provide detailed suggestions.
Mastin et al.\ \cite{mastin2020optimising} examine the spread of plant pathogens. Their paper is close in spirit to our work, in that they focus of optimization of surveillance nodes by building a stochastic optimization model, using Monte Carlo samples. However, their model applies to a ``static'' surveillance strategy in which a fixed number of nodes are selected to aid detection. In this paper we investigate dynamic testing, in which different nodes are tested each day. 

Overall, we note that most previous research focuses on the effectiveness of various surveillance strategies to detect an outbreak of COVID-19, but few papers provide guidance on specific testing strategies, especially when testing resources are in shortage. The model in this paper is predicated on the belief that decisions makers have different risk tolerances, which in turn determine the amount of testing that is performed. We also assume the contacts of individuals in a facility can be heterogeneous. We simulate disease progression on a network and find the probability of detecting an outbreak for various levels of testing capacities and time thresholds, as described in the next section.  
%The contact patterns for individuals can be either obtained from their GPS, Bluetooth, and WiFi data in their cellphone or by a contact tracing questionnaire. 
%Different from Smith et al.\ \cite{Smith2020best} who simulate an initial outbreak scenario as one infection of admitted patient, one staff been infected and one admitted patient been infected outside of the LTC weekly, we assume a homogeneous rate of an individual get infected outside of LTC and randomly generate the initial outbreak condition using a truncated binomial distribution.

\section{Model Description}\label{sec:model}
The first component of our model is a contact graph $G = (V, E)$ on a set $V$ of nodes and a 
set $E$ of undirected edges. We call two nodes \emph{neighbors} if they are connected by an edge. 
Each node in the graph corresponds to a single staff member, and there is an edge connecting them
if they have close contact during a day, i.e., there is a non-zero probability of COVID-19 passing between
them if one of them is infectious. Apart from the graph structure just described, the model is characterized by
the following parameters:
\begin{itemize}
\item $K := |V|$, the number of staff in the facility
\item $p$: daily probability of external infection
\item $\ell$: latency, the number of days to move from the exposed state to the infected state 
\item $d$: degree of each node in the symmetric case, indicating the density of contact graph
\item $r$: daily probability of infection between neighbors, a representation of the force of infection
\item $f$: false negative rate for a disease test, after a node is infectious
\item $t$: outbreak threshold tolerance, in days.
\end{itemize}

We now describe the progression of the disease through the network. The spread model is a discrete-time Markov chain, with time index $n = 0, 1, 2, \ldots.$
At each time point the nodes can be in one of three states: susceptible ($S$), exposed ($E$), and infected ($I$). We do not model recovered nodes since
our focus is on detecting initial outbreaks, generally in less than 10 days from the first infected staff member entering the facility. Nodes may be exposed either
from external interactions, or internal interactions. We also have simplified assumptions regarding the exposed and infected states. In the exposed state
a node cannot spread the disease to other nodes and cannot be detected via a test. After exactly $\ell$ days, an exposed node becomes infected. When
a node is infected, each neighboring susceptible node enters the exposed state with probability $r$ each day. The sequence of events corresponding to infected nodes 
infecting susceptible nodes is assumed to be mutually independent, both among nodes and across days. In addition, to model external infections, 
each susceptible node can move directly to the infected state, each day, with probability $p$. The reason we assume this direct move is that an exposed
node cannot be detected by a test, and cannot infect other nodes, until it is in the ``infected'' state. Hence, from a dynamic systems point of view, there is no
need to track or model nodes that are just ``exposed'' externally. The sequence of events corresponding to external infections are also mutually independent. 
Finally, an external infection resulting in a change to the infected state obviously supersedes a pending move to ``infected'' if the same node has already been
exposed internally. 

In terms of disease testing, a decision maker can administer $k$ tests per day to try to detect an outbreak. For simplicity, we assume that such a test
is administered in the ``morning'' to each of the $k$ individuals chosen that day, and that it is a rapid test, whose results are available before the staff
member interacts with the rest of the staff. From the point of view of our model, this means that the tested individuals are identified immediately if: (a) they
are tested on a given day, (b) are in the infected state on the same day, and (c) the test does not yield a false negative result on this day. At this point, we stop the process, since a detection has occurred. We do not assume that the
test is perfect. Specifically, there is a false negative rate given by $f$. This rate applies to a test given to an individual that is already in the infected state. 
We assume the false negative rate is constant throughout the period a node is infected. This is obviously a simplification of the case for most diseases, 
such as COVID-19, in which the false negative rate fluctuates as an individual moves through the progression of the disease. However, recall that our model
is generally concerned with rapid detection, within the first several days of infection (after the ``incubation'' period given by $\ell$). 

Up until Section \ref{sec:real}, we assume that the contact network is homogeneous in several ways. First, all nodes have an equal
probability of external infection (characterized by $p$). Second, all nodes have equal degree and are vertex transitive (see Section \ref{sec:structure}). Third, each edge
has an equal ``weight,'' in that the infection force $r$ is the same for every edge. In Section \ref{sec:real}, we relax some of these assumptions to model
heterogeneous contact networks, based on more realistic contact data. Part of the reason to first study homogeneous networks is to gain insights into different principles, such as the 
effect of the structure of the graph on testing thresholds and efficient testing protocols. To the best of our knowledge, these issues have not been studied before in the
literature on surveillance testing. 

\subsection{Graph Structure} \label{sec:structure}

In our study of so-called homogeneous graphs, we confine our analysis to $d$-regular graphs, in order to preserve some sense of internal symmetry. 
In addition, this allows us to systematically explore the effect of internal interaction density, as embodied by the single parameter $d$. Of course, 
there are numerous other ways to quantify graph structure, even for $d$-regular graphs, but we do not explore those other avenues in this paper. 

First,
we recall that in a $d$-regular graph each node has exactly $d$ neighbors. A complete graph is a $d$-regular graph in which each node is connected
to every other node in the graph. Although there are many types of $d$-regular graphs for a given $d$, in this paper we only study the class of \emph{circulant
graphs}, which are always $d$-regular. A circulant graph is a graph whose adjacency matrix is a circulant matrix. More specifically, a circulant graph can be
characterized by a vector $(a_1, a_2, \ldots, a_m)$ where $a_1 < a_2 < \cdots < a_{m-1} < a_m$ and $a_m < (K+1)/2$. 
For simplicity, we assume that $a_m < K/2$. In this case, the corresponding circulant graph is $d$-regular with $d = 2m$. 
This allows us to uniquely define $m$, for a specified even $d$ (which we vary to study the effect of graph density on disease spread). 
The vector $(a_1, a_2, \ldots, a_m)$ is sometimes called the \emph{jump sequence}, and it characterizes the connections in the
graph as follows. Suppose we label the nodes in the graph $0, 1, 2, \ldots, K-1$. Then each node $i$ has nodes $(i \pm a_1)  \mod{K}$,
$(i \pm a_2)  \mod{K}$, \ldots, $(i \pm a_m) \mod{K}$, as neighbors. 

\begin{figure}
\begin{centering} 
\includegraphics[width=0.95\textwidth]{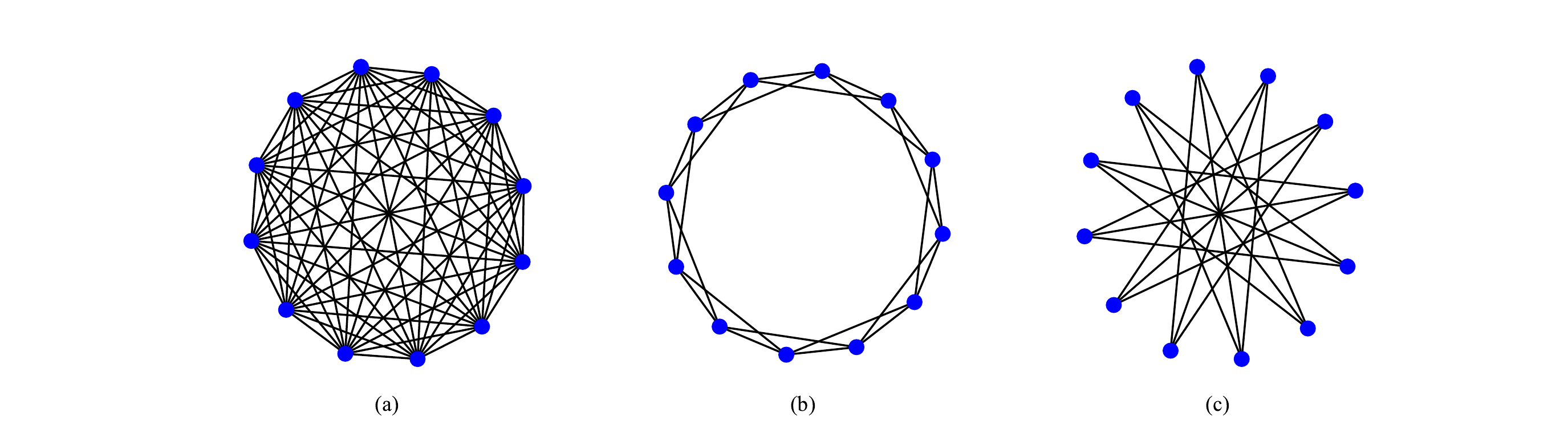}\\
\caption{Some examples of $d$-regular graphs: (a) a complete 11-regular graph, (b) a neighboring 4-regular graph, (c) a crossing 3-regular graph.} \label{fig:regular}
\end{centering}
\end{figure}

Circulant graphs are useful for our study as they have a number of desirable properties. As mentioned above, they are always $d$-regular, with
a degree directly related to the parameter $m$. We also design our graphs so that they are always connected. Further, circulant graphs are \emph{vertex transitive} so that, roughly speaking, every vertex looks similar. Note that circulant graphs need not be edge transitive and hence are not symmetric in the graph-theoretic sense. In studying the effect of graph structure and density on surveillance protocols, we use three classes
of circulant graphs: complete graphs, neighboring graphs, and crossing graphs. The first class is standard. The latter two terms were created for this study. A \emph{neighboring graph} is one in which the jump sequence is given by $(1,2, \ldots, m)$. In other words, each node is connected to its $m$ closest neighbors, if one envisions the nodes being arranged, as labeled, in a circle. A crossing graph is one in which the jump sequence
is given by $( \lfloor K/2-1 \rfloor -m+1, \ldots, \lfloor K/2-1 \rfloor)$. In this case, each is node is connected to its $m$ farthest neighbors. In Figure \ref{fig:regular} we display representatives of each graph class for a small network. 

\subsection{Monte Carlo Simulation} 
In order to answer the questions posed in the introduction, we perform Monte Carlo simulations of disease spread, and detection, on networks
of varying structures and sizes. To keep the insights manageable, the parameters $\ell =3$, $r=0.05$, and $f =0.21$ are kept constant throughout the numerical studies. Otherwise, we vary network size, density, external infection rate, and the outbreak threshold tolerance. 
Given a particular network and testing protocol we typically simulate 50,000 outbreaks in a facility. An outbreak is initiated by at least one
node becoming infectious at time 0, and the resulting nodes that are exposed and infected, until the outbreak threshold tolerance $t$, at which
point the simulation is stopped. For each sample outbreak, we record whether or not there was a successful detection. An outbreak may
not be detected due to an infected node not being tested, or due to a false negative result. These simulations then produce point estimates
and confidence intervals on the true detection probability for a particular testing protocol. In general, we found 50,000 sample paths sufficient
to distinguish among policies. 

In our numerical results, the primary performance criterion is the probability of successfully detecting an outbreak within $t$ days, given
that an outbreak occurs. It is important to note that this is what one might call a conditional outbreak probability. We believe that it is
more useful than computing the unconditional probability of detecting an outbreak, as this type of performance metric essentially penalizes
a protocol for failing to detect an outbreak that does not exist. As we shall see below, this needs to be kept in mind, as some results
under this metric might appear counterintuitive. For example, the probability of successful detection actually increases with the community
infection rate of $p$. To see why, imagine the extreme case in which everyone walks in the door on day 1 infected with a disease and the
false negative rate is 0. In this case, it does not matter who is tested, as everyone will set off the detection alarm. In particular you only
need one test to detect an outbreak when $p=1$ and $f=0$. However, when the
community infection rate is very low, this means that it is very likely that one, and only one, person is the genesis of an outbreak. This makes
the outbreak more difficult to detect with a small number of tests. 

\section{Computational Results for Testing Protocols}

\subsection{Parameter Comparisons} \label{sec:par}

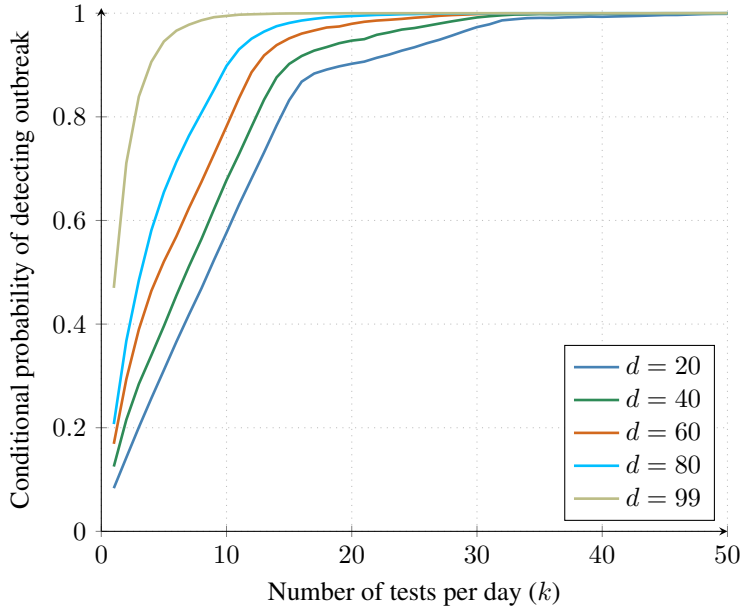
\begin{figure}[ptbh]
    \centering
    \begin{tikzpicture}
        \begin{axis}[axis lines=left, xlabel=Number of tests per day ($k$), ylabel=Conditional probability of detecting outbreak, width=0.6\textwidth, ymajorgrids=true, xmajorgrids=true, grid style=dotted, legend style={at={(0.98, .02)}, anchor=south east, legend cell align=left}, xtick={0, 10, 20, 30, 40, 50}, xmin=0, ytick={0, 0.2, 0.4, 0.6, 0.8, 1}, ymin=0, ymax=1.01]
        \addplot[color=steelblue,line width=1pt] table[x=k, y=deg20_prob, col sep=comma] {filecontents/neighboring_K100_p0.0001_r0.05_f0.21_l3_t6_varyingdegree.csv};
        \addlegendentry{$d=20$};
        \addplot[color=seagreen,line width=1pt] table[x=k, y=deg40_prob, col sep=comma] {filecontents/neighboring_K100_p0.0001_r0.05_f0.21_l3_t6_varyingdegree.csv};
        \addlegendentry{$d=40$};
        \addplot[color=chocolate,line width=1pt] table[x=k, y=deg60_prob, col sep=comma] {filecontents/neighboring_K100_p0.0001_r0.05_f0.21_l3_t6_varyingdegree.csv};
        \addlegendentry{$d=60$};
        \addplot[color=deepskyblue,line width=1pt] table[x=k, y=deg80_prob, col sep=comma] {filecontents/neighboring_K100_p0.0001_r0.05_f0.21_l3_t6_varyingdegree.csv};
        \addlegendentry{$d=80$};
        \addplot[color=tabolive,line width=1pt] table[x=k, y=deg99_prob, col sep=comma] {filecontents/neighboring_K100_p0.0001_r0.05_f0.21_l3_t6_varyingdegree.csv};
        \addlegendentry{$d=99$};
        \end{axis}
    \end{tikzpicture}
    \caption{Number of tests versus detection probability for circulant neighboring graphs of varying degrees. Fixed graph parameters are $K=100$, $p=0.0001$, $r=0.05$, $f=0.21$, $\ell=3$, $t=6$. The results were obtained by sampling 50,000 virus sample paths.}
    \label{fig:degree}
\end{figure}

In our first set of experiments, we investigate the effect of graph density, embodied by the parameter $d$, on the number of tests
per day needed to detect an outbreak. In our model, $d$ represents the number of other staff members a particular staff member is
in contact with each day. For example, when $K =100$ and $d=99$, the resulting contact network is a complete graph, and the
graph implies that all staff members come into contact with all other staff members each day. Obviously, this is an extreme situation. 
As such, we also examine lower density networks. In Figure \ref{fig:degree}, we examine the case of 100 staff members and an external 
community infection probability of $p=0.0001$ (per day). We set $t=6$, indicating that our goal is to detect an outbreak within 6 days
of the introduction of the disease into the facility. Keeping these parameters fixed, we plot the point estimates for the probability of 
detection versus the number of tests per day, and plot this curve for various values of $d$. The first thing to note is that there appear to be diminishing
marginal benefits of administering more tests, as one might expect. From the graph we also see that the number of tests required
to meet a certain benchmark probability of detection can vary significantly with $d$. For example, if our threshold probability is 0.7, and $d=99$
then roughly \textcolor{black}{4} tests per day are needed. However, if $d=20$ then about \textcolor{black}{13} tests per day are needed. This may seem counterintuitive but
it is related to our observation above relating to the community infection rate. The more quickly a disease spreads in a facility, the more
quickly it can be detected. We also observe that if we wish to enforce a very high (say 0.95) detection probability, then a large number of 
staff need to be tested daily. This is unlikely to be palatable for management and staff. 

%\begin{figure}
%\begin{centering} 
%\includegraphics[width=15cm]{degree.jpg}\\
%\caption{Number of tests versus detection probability for circulant neighboring graphs of varying degrees. Fixed graph parameters are $K=100$, $p=0.0001$, $r=0.05$, $f=0.21$, $\ell=3$, $t=6$. The results were obtained by sampling 50,000 virus sample paths.} \label{fig:degree}
%\end{centering}
%\end{figure}

% https://tex.stackexchange.com/questions/215989/controlling-pfplots-axis-labels

\begin{figure}[ptbh]
    \centering
    \begin{tikzpicture}
        \begin{axis}[axis lines=left, xlabel=Number of tests per day ($k$), ylabel=Conditional probability of detecting outbreak, width=0.6\textwidth, ymajorgrids=true, xmajorgrids=true, grid style=dotted, legend style={at={(0.98, .02)}, anchor=south east, legend cell align=left}, xtick={0, 10, 20, 30, 40, 50}, xmin=0, ytick={0, 0.2, 0.4, 0.6, 0.8, 1}, ymin=0, ymax=1.01]
        \addplot[color=steelblue,line width=1pt] table[x=k, y=tol4_prob, col sep=comma] {filecontents/neighboring_K100_p0.01_r0.05_f0.21_l3_d60_varyingt.csv};
        \addlegendentry{$t=4$};
        \addplot[color=seagreen,line width=1pt] table[x=k, y=tol5_prob, col sep=comma] {filecontents/neighboring_K100_p0.01_r0.05_f0.21_l3_d60_varyingt.csv};
        \addlegendentry{$t=5$};
        \addplot[color=chocolate,line width=1pt] table[x=k, y=tol6_prob, col sep=comma] {filecontents/neighboring_K100_p0.01_r0.05_f0.21_l3_d60_varyingt.csv};
        \addlegendentry{$t=6$};
        \addplot[color=deepskyblue,line width=1pt] table[x=k, y=tol7_prob, col sep=comma] {filecontents/neighboring_K100_p0.01_r0.05_f0.21_l3_d60_varyingt.csv};
        \addlegendentry{$t=7$};
        \addplot[color=tabolive,line width=1pt] table[x=k, y=tol8_prob, col sep=comma] {filecontents/neighboring_K100_p0.01_r0.05_f0.21_l3_d60_varyingt.csv};
        \addlegendentry{$t=8$};
        \addplot[color=taborange,line width=1pt] table[x=k, y=tol9_prob, col sep=comma] {filecontents/neighboring_K100_p0.01_r0.05_f0.21_l3_d60_varyingt.csv};
        \addlegendentry{$t=9$};
        \addplot[color=tabpink,line width=1pt] table[x=k, y=tol10_prob, col sep=comma] {filecontents/neighboring_K100_p0.01_r0.05_f0.21_l3_d60_varyingt.csv};
        \addlegendentry{$t=10$};
        \end{axis}
    \end{tikzpicture}
    \caption{Number of tests versus detection probability for circulant neighboring graphs of tolerance levels. Fixed graph parameters are $K=100$, $p=0.01$, $r=0.05$, $f=0.21$, $\ell=3$, $d=60$. The results were obtained by sampling 50,000 virus sample paths.}
    \label{fig:tolerance}
\end{figure}
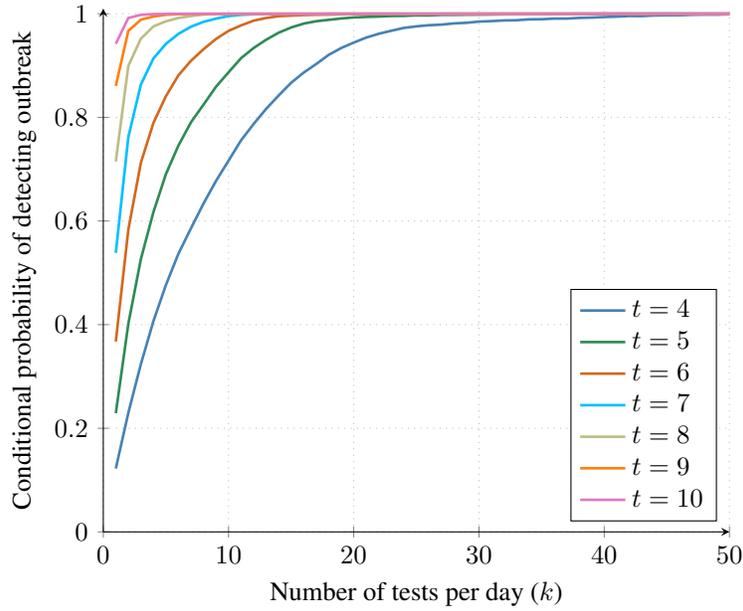

In the next set of experiments, we vary the outbreak threshold tolerance $t$ while keeping other parameters fixed. We choose a neighboring graph with $d=60$ as the results for this graph show the contrast among different tolerance levels well. The results are shown in Figure \ref{fig:tolerance}. As must be the case, the detection probability curves are nested, with larger tolerance curves dominating lower tolerance curves. Again we notice that the number of tests required to detect an outbreak at various tolerance levels varies widely. If the probability threshold is again 0.7, and $t=8$ then just \textcolor{black}{one} test per day is needed. However, if $t=4$ then around \textcolor{black}{10} tests per day are needed, which implies that all staff are tested every \textcolor{black}{10} days. %Another feature to notice is that the graph for $t=4$ is nearly linear. This is to be expected because when $\ell=3$ the network effects of virus spread have a minimal effect on the dynamics of virus detection. 

%\begin{figure} 
%\begin{centering}
%%\includegraphics[width=15cm]{tolerance.jpg}\\
%\includegraphics[width=15cm]{external.jpg}\\
%\caption{Number of tests versus detection probability for circulant neighboring graphs of tolerance levels. Fixed graph parameters are $K=100$, $p=0.01$, $r=0.05$, $f=0.21$, $\ell=3$, $d=60$. The results were obtained by sampling 50,000 virus sample paths.}  \label{fig:tolerance}
%\end{centering}
%\end{figure}

%% https://tex.stackexchange.com/questions/215989/controlling-pfplots-axis-labels

\begin{figure}[ptbh]
    \centering
    \begin{tikzpicture}
        \begin{axis}[axis lines=left, xlabel=Number of tests per day ($k$), ylabel=Conditional probability of detecting outbreak, width=0.6\textwidth, ymajorgrids=true, xmajorgrids=true, grid style=dotted, legend style={at={(0.98, .02)}, anchor=south east, legend cell align=left}, xtick={0, 10, 20, 30, 40, 50}, xmin=0, ytick={0, 0.2, 0.4, 0.6, 0.8, 1}, ymin=0, ymax=1.01]
        \addplot[color=steelblue,line width=1pt] table[x=k, y=tol4_prob, col sep=comma] {filecontents/neighboring_K100_p0.0001_r0.05_f0.21_l3_d60_varyingt.csv};
        \addlegendentry{$t=4$};
        \addplot[color=seagreen,line width=1pt] table[x=k, y=tol5_prob, col sep=comma] {filecontents/neighboring_K100_p0.0001_r0.05_f0.21_l3_d60_varyingt.csv};
        \addlegendentry{$t=5$};
        \addplot[color=chocolate,line width=1pt] table[x=k, y=tol6_prob, col sep=comma] {filecontents/neighboring_K100_p0.0001_r0.05_f0.21_l3_d60_varyingt.csv};
        \addlegendentry{$t=6$};
        \addplot[color=deepskyblue,line width=1pt] table[x=k, y=tol7_prob, col sep=comma] {filecontents/neighboring_K100_p0.0001_r0.05_f0.21_l3_d60_varyingt.csv};
        \addlegendentry{$t=7$};
        \addplot[color=tabolive,line width=1pt] table[x=k, y=tol8_prob, col sep=comma] {filecontents/neighboring_K100_p0.0001_r0.05_f0.21_l3_d60_varyingt.csv};
        \addlegendentry{$t=8$};
        \addplot[color=taborange,line width=1pt] table[x=k, y=tol9_prob, col sep=comma] {filecontents/neighboring_K100_p0.0001_r0.05_f0.21_l3_d60_varyingt.csv};
        \addlegendentry{$t=9$};
        \addplot[color=tabpink,line width=1pt] table[x=k, y=tol10_prob, col sep=comma] {filecontents/neighboring_K100_p0.0001_r0.05_f0.21_l3_d60_varyingt.csv};
        \addlegendentry{$t=10$};
        \end{axis}
    \end{tikzpicture}
    \caption{Number of tests versus detection probability for circulant neighboring graphs of varying tolerance levels. Fixed graph parameters are $K=100$, $p=0.0001$, $r=0.05$, $f=0.21$, $\ell=3$, $d=60$. The results were obtained by sampling 50,000 virus sample paths.}
    \label{fig:external}
\end{figure}
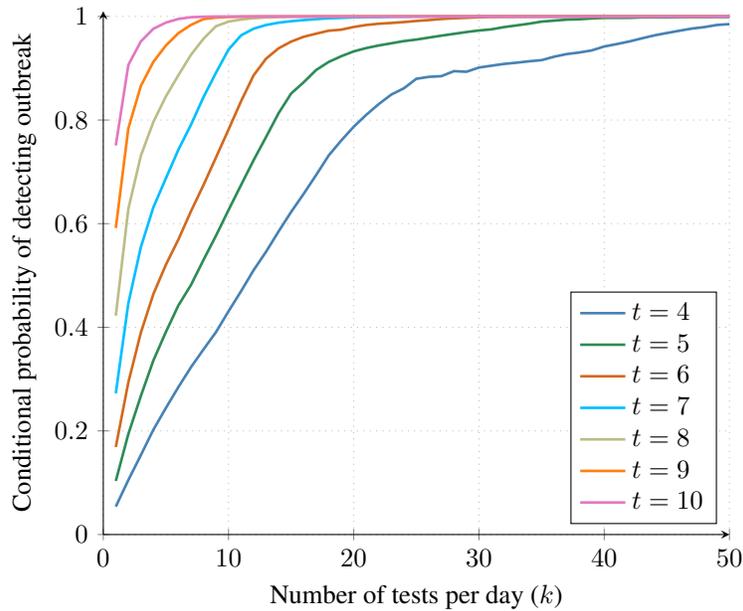

Finally, we examine the effect of the external infection probability on the detection probability curves. The networks tested
in Figure \ref{fig:external} have the same parameters as the networks tested in Figure \ref{fig:tolerance}, with one exception: we change the value of $p$ from 0.01 to $0.0001$.  For each tolerance level $t$ we note that the curve corresponding to 
$p=0.01$ dominates the curve for $p=0.0001$. For example when the probability threshold is 0.7, $t=4$, and $p=0.01$, 
around \textcolor{black}{10} tests per day are needed, as observed above. However, when we examine the case with $p=0.0001$ the required
number of tests per day is approximately \textcolor{black}{17}. At first glance, this may seem counterintuitive: we require fewer tests when a disease is more prevalent in the community. However, the result is mathematically sound. For example, suppose we examine the extreme case when $p$ is very close to 1. In that case, a large number of staff members would arrive to the facility each day being infectious. As such, we only need to test a small fraction of the staff to detect an outbreak. In contrast, when
the community infection rate is low, it is likely that only 1 staff member out of, say 100, would initiate an outbreak. This one
seed of an outbreak is harder to catch with a low number of tests.

%\begin{figure} 
%\begin{centering}
%%\includegraphics[width=12cm]{external.jpg}\\
%\includegraphics[width=12cm]{tolerance.jpg}\\
%\caption{Number of tests versus detection probability for circulant neighboring graphs of varying tolerance levels. Fixed graph parameters are $K=100$, $p=0.0001$, $r=0.05$, $f=0.21$, $\ell=3$, $d=60$. The results were obtained by sampling 50,000 virus sample paths.}  \label{fig:external}
%\end{centering}
%\end{figure}

%% https://tex.stackexchange.com/questions/215989/controlling-pfplots-axis-labels

\subsection{The Effect of Testing Order} \label{sec:testo}
In the experiments of Section \ref{sec:par}, the nodes of the graph were tested in ``standard order.'' What this means
is that we envision the nodes with a fixed numbering, 0 to $K-1$, and arranged in a circle according to the numbering. 
This numbering is used to characterize various types of circulant graphs, as described already. In addition, the standard
testing order is to begin by testing nodes 0 through $k-1$ on day 1, then testing nodes $k$ through $2k-1$ on day 2, etc.
We continue testing in this manner, returning to node 1 when all nodes have been tested. Of course, for some values
of $t$ and $K$ not all nodes will be tested. With respect to our simulation results, it does not matter with which node we start since outbreaks are simulated in a uniform manner across all nodes, and the graphs are vertex transitive. However, it is not clear that testing in this standard order, for any given initial starting node, is optimal. Obtaining the optimal testing order is computationally prohibitive, as the underlying optimization problem is a stochastic integer program, with a very large number of feasible solutions. In this paper, we make no attempt to prove optimality bounds on testing orders, or to develop complex heuristics. Rather, we examine a few different testing algorithms on circulant graphs. Our results indicate that the testing order has little effect on the detection probability, at least for circulant graphs. This is good news in that it indicates that optimizing the testing order likely is not a high priority in tuning test protocols in contact networks. 

\begin{figure}[ptbh]
    \centering
    \begin{tikzpicture}
        \begin{axis}[axis lines=left, xlabel=Number of tests per day ($k$), ylabel=Conditional probability of detecting outbreak, width=0.6\textwidth, ymajorgrids=true, xmajorgrids=true, grid style=dotted, legend style={at={(0.98, .02)}, anchor=south east, legend cell align=left}, xtick={0, 2, 4, 6, 8, 10, 12, 14, 16, 18, 20}, xmin=0, ytick={0, 0.2, 0.4, 0.6, 0.8, 1}, ymin=0, ymax=1.01]
        \addplot[color=taborange,line width=1pt] table[x=k, y=circular_prob, col sep=comma] {filecontents/neighboring_K100_p0.01_r0.05_f0.21_l3_d80_t6_varyingtestprotocol.csv};
        \addlegendentry{circle};
        \addplot[color=deepskyblue,line width=1pt] table[x=k, y=random_prob, col sep=comma] {filecontents/neighboring_K100_p0.01_r0.05_f0.21_l3_d80_t6_varyingtestprotocol.csv};
        \addlegendentry{random};
        \addplot[color=csspurple,line width=1pt] table[x=k, y=new_prob, col sep=comma] {filecontents/neighboring_K100_p0.01_r0.05_f0.21_l3_d80_t6_varyingtestprotocol.csv};
        \addlegendentry{new};
        \end{axis}
    \end{tikzpicture}
    \caption{Number of tests versus detection probability for various testing protocols in a neighboring graph. Fixed graph parameters are $K=100$, $p=0.01$, $r=0.05$, $f=0.21$, $\ell=3$, $d=80$, $t=6$. The results were obtained by sampling 10,000 virus sample paths.}
    \label{fig:order1}
\end{figure}
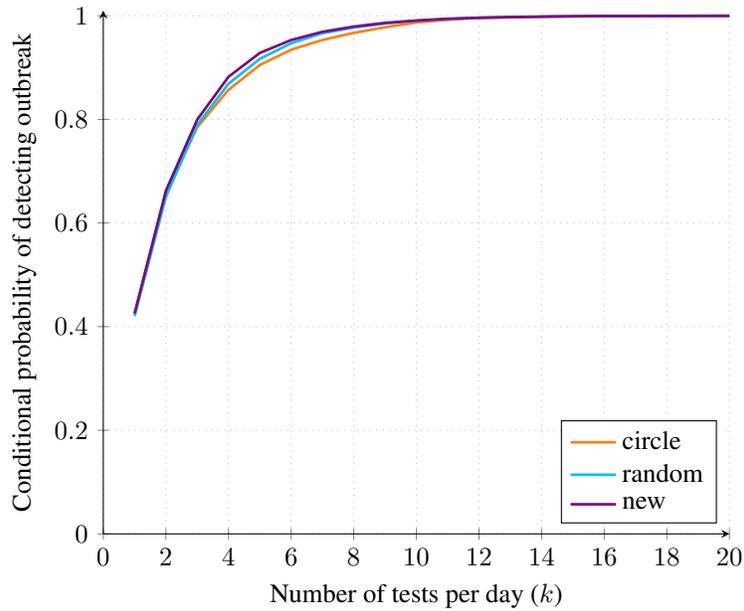

In our next set of experiments, we examine the effect of changing the testing order, while keeping all the network
parameters fixed. In Figure \ref{fig:order1} we examine three possible testing protocols. The first protocol, labeled ``circle''
is the standard order described in the previous paragraph. The second protocol, labeled ``new'' works as follows. The first node to be tested is chosen uniformly at random. Call this node $A$. The second node that is tested is the node that has the maximum distance from $A$, with ties broken randomly. Recall that the distance between two nodes in a graph is the number of edges in the shortest path connecting the nodes. To continue the selection process after some
set of nodes has already been selected, the next node chosen is one
whose sum of distances from all previously selected nodes is largest. 
To break ties among equal distance sums, we select a node for which these distances have the smallest sample standard deviation (if there is still a tie, it is broken randomly).  We repeat the selection process until all nodes have been selected, to create an ordered list that defines the testing order. In our simulation analysis, we test the nodes in the given order and then return to the beginning of the list once all nodes have been tested once. The third protocol is based on a randomly selected permutation of the
numbers 0 through $K-1$. 

For the neighboring graph testing in Figure \ref{fig:order1}, we see that the distance algorithm
does slightly outperform the circle protocol for testing levels of about 5 to 10 nodes per day. This also implies a slight
difference in the required number of tests at the mid-level probability thresholds. For example, for a probability threshold
of 0.9 the difference in the required tests per day appears to be about 2. Outside this middle zone, the difference is essentially negligible. Note that we do expect the circle protocol to have a slightly lower performance in a neighboring graph. When the virus is most likely to spread among close neighbors, when the nodes are arranged in a circle, then there is some redundancy in testing nodes in this same circular order. It is better to hop across the circle for subsequent tests, which is essentially what the ``new'' protocol does.

%\begin{figure} 
%\begin{centering}
%\includegraphics[width=12cm]{neighbouring.png}\\
%\caption{Number of tests versus detection probability for various testing protocols in a neighboring graph. Fixed graph parameters are $K=100$, $p=0.01$, $r=0.05$, $f=0.21$, $\ell=3$, $d=80$, $t=6$. The results were obtained by sampling 10,000 virus sample paths.}  \label{fig:order1}
%\end{centering}
%\end{figure}

%% https://tex.stackexchange.com/questions/215989/controlling-pfplots-axis-labels

\begin{figure}[ptbh]
    \centering
    \begin{tikzpicture}
        \begin{axis}[axis lines=left, xlabel=Number of tests per day ($k$), ylabel=Conditional probability of detecting outbreak, width=0.6\textwidth, ymajorgrids=true, xmajorgrids=true, grid style=dotted, legend style={at={(0.98, .02)}, anchor=south east, legend cell align=left}, xtick={0, 2, 4, 6, 8, 10, 12, 14, 16, 18, 20}, xmin=0, ytick={0, 0.2, 0.4, 0.6, 0.8, 1}, ymin=0, ymax=1.01]
        \addplot[color=taborange,line width=1pt] table[x=k, y=circular_prob, col sep=comma] {filecontents/crossing_K100_p0.01_r0.05_f0.21_l3_d80_t6_varyingtestprotocol.csv};
        \addlegendentry{circle};
        \addplot[color=deepskyblue,line width=1pt] table[x=k, y=random_prob, col sep=comma] {filecontents/crossing_K100_p0.01_r0.05_f0.21_l3_d80_t6_varyingtestprotocol.csv};
        \addlegendentry{random};
        \addplot[color=csspurple,line width=1pt] table[x=k, y=new_prob, col sep=comma] {filecontents/crossing_K100_p0.01_r0.05_f0.21_l3_d80_t6_varyingtestprotocol.csv};
        \addlegendentry{new};
        \end{axis}
    \end{tikzpicture}
    \caption{Number of tests versus detection probability for various testing protocols in a crossing graph. Fixed graph parameters are $K=100$, $p=0.01$, $r=0.05$, $f=0.21$, $\ell=3$, $d=80$, $t=6$. The results were obtained by sampling 50,000 virus sample paths.}
    \label{fig:order2}
\end{figure}
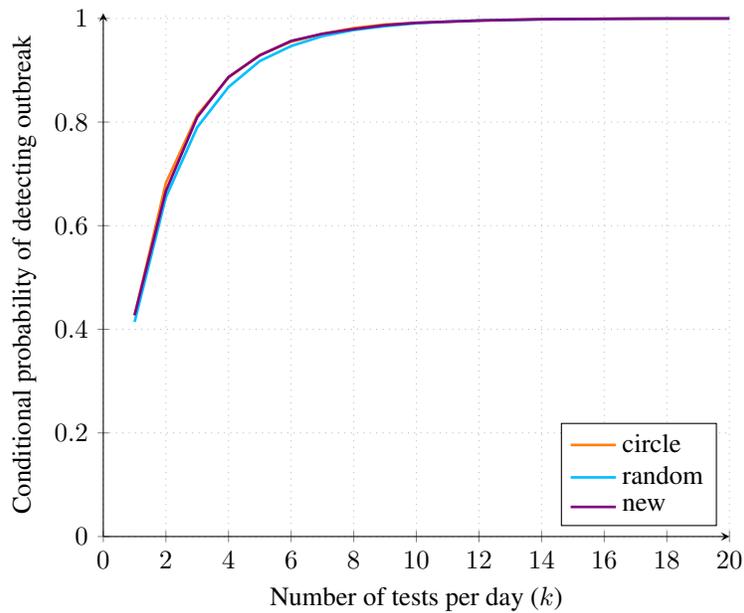

We next test the same three testing protocols in a crossing graph. In Figure \ref{fig:order2}, we see that there is almost no difference in the three protocols, across a range of testing levels. This is also expected in this case. In a crossing graph, the testing protocols circle and new are quite similar, because nodes that are ``neighbors'' when arranged on a circle are actually rather distant as measured by the number of hops required to travel between such nodes. Again, these results indicate that testing protocol does not have a large effect on detection probability, at least for circulant graphs. 

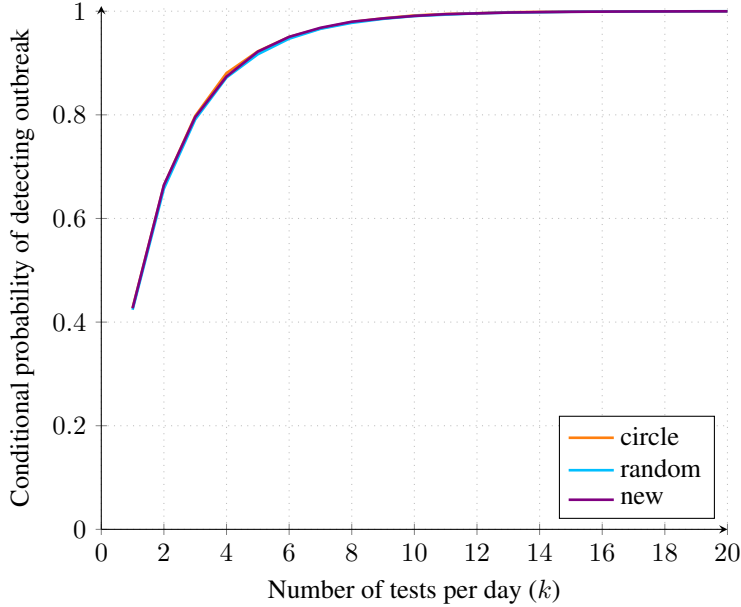
\begin{figure}[ptbh]
    \centering
    \begin{tikzpicture}
        \begin{axis}[axis lines=left, xlabel=Number of tests per day ($k$), ylabel=Conditional probability of detecting outbreak, width=0.6\textwidth, ymajorgrids=true, xmajorgrids=true, grid style=dotted, legend style={at={(0.98, .02)}, anchor=south east, legend cell align=left}, xtick={0, 2, 4, 6, 8, 10, 12, 14, 16, 18, 20}, xmin=0, ytick={0, 0.2, 0.4, 0.6, 0.8, 1}, ymin=0, ymax=1.01]
        \addplot[color=taborange,line width=1pt] table[x=k, y=circular_prob, col sep=comma] {filecontents/randomdregular_K100_p0.01_r0.05_f0.21_l3_d80_t6_varyingtestprotocol.csv};
        \addlegendentry{circle};
        \addplot[color=deepskyblue,line width=1pt] table[x=k, y=random_prob, col sep=comma] {filecontents/randomdregular_K100_p0.01_r0.05_f0.21_l3_d80_t6_varyingtestprotocol.csv};
        \addlegendentry{random};
        \addplot[color=csspurple,line width=1pt] table[x=k, y=new_prob, col sep=comma] {filecontents/randomdregular_K100_p0.01_r0.05_f0.21_l3_d80_t6_varyingtestprotocol.csv};
        \addlegendentry{new};
        \end{axis}
    \end{tikzpicture}
    \caption{Number of tests versus detection probability for various testing protocols in a random $d$-regular graph. Fixed graph parameters are $K=100$, $p=0.01$, $r=0.05$, $f=0.21$, $\ell=3$, $d=80$, $t=6$. The results were obtained by sampling 10,000 virus sample paths.}
    \label{fig:order3}
\end{figure}

Finally, in Figure \ref{fig:order3}, we test the same protocols on a randomly generated $d$-regular graph. As expected, there are relatively small differences among protocols. We also tested protocols for different graph densities, and the results are similar to those presented herein. We conclude that the difference among testing orders is usually insignificant, with some minor differences as shown in Figure \ref{fig:order1}. Since the ``new'' protocol performs well in all cases, we recommend this protocol, assuming that the decision maker has concrete knowledge of the graph structure. Otherwise, nodes can likely be tested in an arbitrary manner, with little loss of performance versus a good heuristic. 

%\begin{figure} 
%\begin{centering}
%\includegraphics[width=12cm]{crossing.png}\\
%\caption{Number of tests versus detection probability for various testing protocols in a crossing graph. Fixed graph parameters are $K=100$, $p=0.01$, $r=0.05$, $f=0.21$, $\ell=3$, $d=80$, $t=6$. The results were obtained by sampling 50,000 virus sample paths.}  \label{fig:order2}
%\end{centering}
%\end{figure}

%% https://tex.stackexchange.com/questions/215989/controlling-pfplots-axis-labels

%\begin{figure} 
%\begin{centering}
%\includegraphics[width=12cm]{random.png}\\
%\caption{Number of tests versus detection probability for various testing protocols in a random $d$-regular graph. Fixed graph parameters are $K=100$, $p=0.01$, $r=0.05$, $f=0.21$, $\ell=3$, $d=80$, $t=6$. The results were obtained by sampling 10,000 virus sample paths.}  \label{fig:order3}
%\end{centering}
%\end{figure}

%% https://tex.stackexchange.com/questions/215989/controlling-pfplots-axis-labels

\section{Heterogeneous Graphs and Real-world Test Cases} \label{sec:real}

\subsection{Real-world Test Cases} 
\subsubsection{Model adjustment}
In previous sections, we considered simplified contact networks in order to gain insight into the relationship between the parameters of the graph, and the number of tests needed to detect an outbreak. In the remaining sections, we consider more general graphs that are inspired by real-world situations and data. A primary difference
is that we allow the parameters $r$ and $p$ to be heterogeneous, i.e., they can vary by node. The outside probability of infection $p$ may vary since each staff member has different social interactions and risk outside of work. Also, in real facilities staff in different categories can have very different contact patterns. Hence, the corresponding contact graph is no longer ``symmetric,'' and the internal infection probability $r$ can vary by node. Our assumptions regarding the latency $l$, the false negative rate $f$, and disease testing procedures are the same as in Section \ref{sec:model}.  

\textcolor{black}{In many facilities, the staff rosters differ each day as do the contact patterns. In theory, it might be possible to develop even more nuanced models in a particular facility. For example, Duval et al.\ \cite{duval2018measuring} studied the contact patterns in one nursing home by asking the staff and patients to use wearable sensors and recorded the contact duration, distances, and frequencies. However, for legal and ethical reasons, there are obviously significant barriers to collecting such detailed data in most facilities. Hence, we do not attempt to model a representative facility at this level of detail. Instead, we collect general data on staff levels, categories and interactions. We believe that this level of modeling still provides important insight into appropriate testing levels for outbreak detection.}

 Let
 $c$ be the number of different staff categories in a facility. 
The staff categories are given by the $C=\{1,2,...,c\}$. Since we have heterogeneous internal infection parameters, we now use 
$r_{c_1,c_2}$ to denote the daily probability of infection between staff members in categories $c_1$ and $c_2$, assuming that they are connected by an edge in the contact graph. Of course, we allow $c_1=c_2$ to reflect interactions between staff in the same category. 
Staff in a facility can be categorized by their job titles, rosters, offices, working units, etc. In our real-world case study, we categorize the staff in a nursing home by job title, and we assume the staff with the same job titles share the same contact patterns. 

\subsubsection{Parameter fitting}
We estimate the parameter $r_{c_1,c_2}$ for different choices of $c_1$ and $c_2$ based on the contact patterns among staff. 
Inspired by Smith et al.\ \cite{Smith2020best} and Temime et al.\ \cite{10.1093/cid/ciaa682}, we use the contact frequencies, average duration of each contact, and the rate of infection with close contact
%and the basic reproduction number $R_0$ 
to estimate the probability of internal infection for staff in various categories. 
%A simple formula for the basic reproduction number is $R_0=p_0\cdot d_0\cdot n_0\cdot i_0 $, where 
Let $p_0$ be the probability of infection per minute between two individuals with close contact, $d_0$ the average contact duration (in minutes), and $n_0$ the average contact frequency. 
%and $i_0$ the average duration of infectivity (in days). 
%Therefore, if the reproduction number $R_0$ is known, we can estimate the probability of infection per minute as
%\begin{equation*}
%    p_0=\frac{R_0}{d_0\cdot n_0\cdot i_0 }.
%\end{equation*}
Then we can estimate the internal probability of infection as $ r=p_0\cdot d_0\cdot n_0$.
We collected data from a nursing home in the northern US by collecting information from the facility manager. The details of this contact data are shown in 
Table \ref{tab:raw_data}. Next, for each staff category, we estimated the number of other staff they come into contact with daily, for all other categories. This data is shown in Table \ref{tab:edges_fitting}. For a job category in row $i$ of the table, the $i,j$th entry in the upper part of the table is the average number of contacts for a staff member of category $i$ with staff members in category $j$. Note that the table would not be symmetric if the lower half was filled in, but the data in the upper portion is sufficient to estimate contacts between all pairs of staff categories.

%By talking with the Manager, we find it is hard for the staff to recall exactly the number of staff that they meet in various categories because it can change daily. So, they organized the data as shown in Table \ref{tab:raw_data}.

\begin{table}
\begin{small}
\begin{tabular}{l c c c c c}
\hline
\textbf{Job title} & \multicolumn{1}{m{1.5cm}}{\textbf{Staff \#}} & \multicolumn{1}{p{2.5cm}}{\textbf{\# of daily \newline contact \newline with residents}}  & \multicolumn{1}{p{2.2cm}}{\textbf{\# of daily \newline contact \newline with staff}} & \multicolumn{1}{p{2.2cm}}{\textbf{Avg.\ contact \newline distance \newline (ft)}} & \multicolumn{1}{p{2.2cm}}{\textbf{Avg.\ contact \newline duration \newline (min)}}  \\ \hline
\textbf{Nurses and nurse aids} &69  &30  &20  & $<$ 6  & $>$ 15  \\ \hdashline
\textbf{House keeping} &16 & 10  &10  &$>$ 6  &$<$ 15   \\ \hdashline
\textbf{Rehabilitation} &4 & 10 & 10 & $<$ 6 & $>$ 15 \\ \hdashline
\textbf{Activity lifestyle} &12 & 20 & 10 &$<$ 6  &15  \\ \hdashline
\textbf{Salon}&2 & 10  & 10 & $<$ 6 & 15  \\ \hdashline
\textbf{Maintenance}&12 &10  & 10 & $<$ 6  & $<$ 15  \\ \hdashline
\textbf{Reception}&6 & 10 & 40 & $<$ 6  & $>$ 15 \\ \hdashline
\textbf{Dining}&6 & 30 & 20  & $<$ 6 & $<$ 15 \\ \hdashline
\textbf{Social}&5 & 10  & 10 & $<$ 6 & $<$ 15 \\ \hdashline
\textbf{Finance}&14 & 0 & 10 & $<$ 6 & $>$ 15  \\ \hdashline
\textbf{Logistics}&3 & 0  & 20 & $<$ 6 & $<$ 15  \\ \hdashline
\textbf{Driver}&4 & 20 & 10 & $<$ 6 & $>$ 15 \\ \hline
\end{tabular}
\end{small}
\caption{Staff data for the nursing home case study} \label{tab:raw_data}\centering
\end{table}

%As shown in Table \ref{tab:edges_fitting}, we can see that the upper triangle of the table embodies enough information of the connected edges. Also, since the values in the table is an estimated averaged number, the average numbers of staff the given staff meets shown on the table can be decimal.
\begin{table}\small
    \centering
    \begin{tabular}{ccccccccccccc}
        \hline
        \textbf{Job title}& \textbf{NR} & \textbf{HK} & \textbf{RH} & \textbf{MS} & \textbf{SL} & \textbf{MT} &\textbf{RC} &\textbf{DN} & \textbf{SC} &\textbf{AD} &\textbf{LG} &\textbf{DR} \\ \hline
        \textbf{NR} &15 &1.5 &0.2 &1 &0.1 &0.1 &2  &0.5 &1 &1 &0.5 &0.22\\ \hdashline
        \textbf{HK} &$-$ &4 &0.5 &0.5 &0.1 &1 &1 &0.2 &0.2 &1 &0.1 &0.25\\ \hdashline
        \textbf{RH} &$-$ &$-$ &1.5 &2 &0.1 &0.1 &0.5 &0.2 &0.2 &1 &0.5 &0.2\\ \hdashline
        \textbf{MS} &$-$ &$-$ &$-$ &3 &0.1 &0.1 &0.5 &0.2 &0.2 &1 &0.5 &0.2\\ \hdashline
        \textbf{SL} &$-$ &$-$ &$-$ &$-$ &1 &0.1 &0.5 &0.1 &0.1 &1.2 &0.5 &0.2\\ \hdashline
        \textbf{MT} &$-$ &$-$ &$-$ &$-$ &$-$ &4 &0.5 &0.1 &0.1 &1.2 &0.5 &0.3\\ \hdashline
        \textbf{RC} &$-$ &$-$ &$-$ &$-$ &$-$ &$-$ &2 &0.1 &0.1 &3 &1 &0.4\\ \hdashline
        \textbf{DN} &$-$ &$-$ &$-$ &$-$ &$-$ &$-$ &$-$ &4 &1 &1 &0.1 &0.1\\ \hdashline
        \textbf{SC} &$-$ &$-$ &$-$ &$-$ &$-$ &$-$ &$-$ &$-$ &3 &1 &0.5 &0.2\\ \hdashline
        \textbf{AD} &$-$ &$-$ &$-$ &$-$ &$-$ &$-$ &$-$ &$-$ &$-$ &5 &1 &0.5\\ \hdashline
        \textbf{LG} &$-$ &$-$ &$-$ &$-$ &$-$ &$-$ &$-$ &$-$ &$-$ &$-$ &1.5 &0.4\\ \hdashline
        \textbf{DR} &$-$ &$-$ &$-$ &$-$ &$-$ &$-$ &$-$ &$-$ &$-$ &$-$ &$-$ &0.5\\ \hline
    \end{tabular}
    \caption{Estimated daily contacts between staff of different categories, rounded to two decimals places}\label{tab:edges_fitting}
\end{table}

\begin{figure} 
\begin{centering}
\includegraphics[width=12cm]{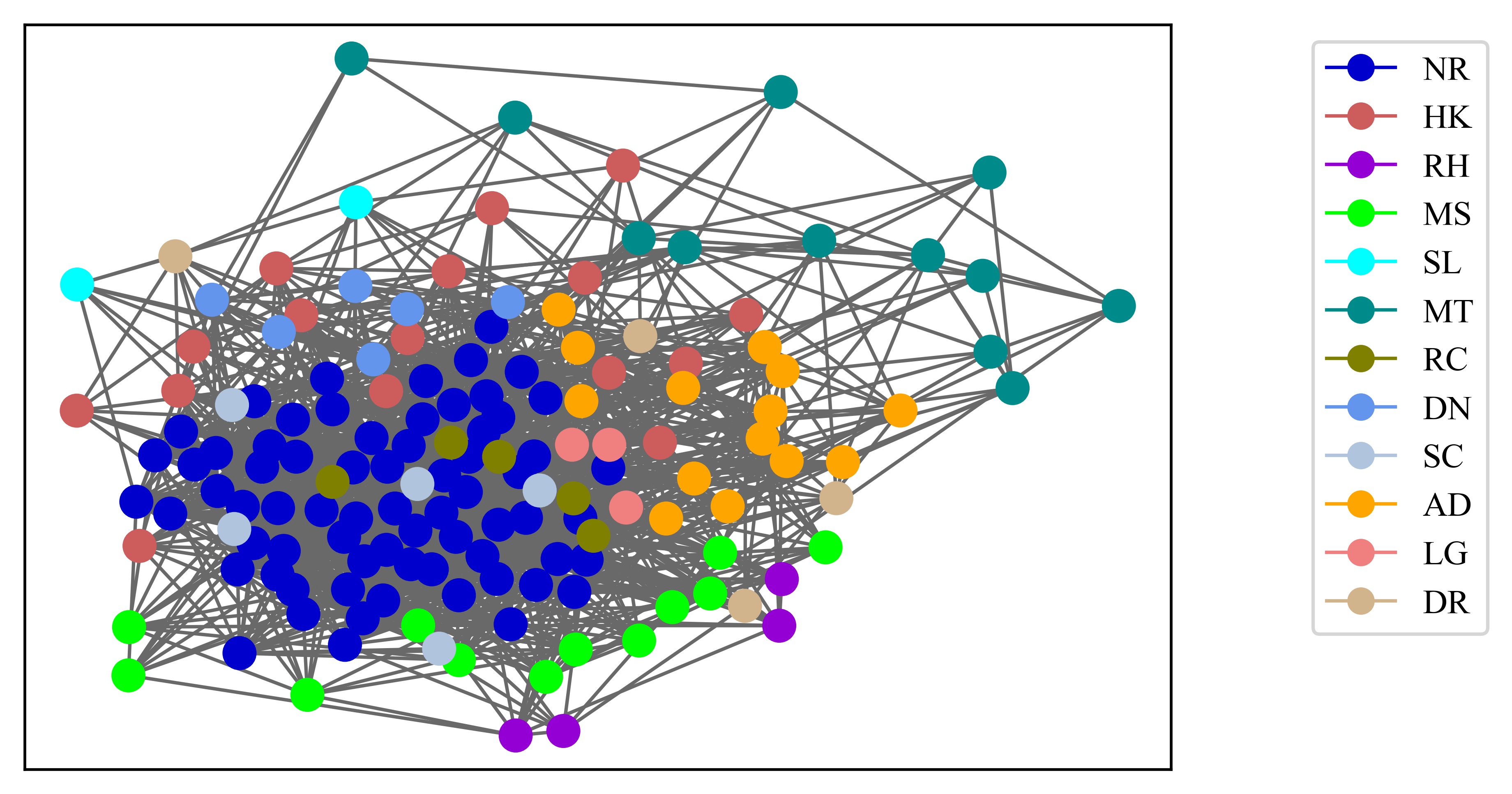}
\caption{A real-world LTC network with $s=152, c=12$}  \label{fig:Example of real-life network}
\end{centering}
\end{figure}

Based on this data, we are now able to generate a representative contact graph for the case-study nursing home. First, we label all the nodes in the graph with their corresponding staff category. Then, we calculate the probability of generating an edge between two nodes by using the data in Tables \ref{tab:raw_data} and \ref{tab:edges_fitting}. For example, if a node is labeled as a nurse, the probability that a contact is triggered with another nurse node is $15/68\approx 0.22$, as each nurse has an average of 15 contacts with the other 68 nurses during a shift.
\textcolor{black}{Then, for every pair of nurses, we generate a Bernoulli(0.22) random variable and assign an edge to this pair if the outcome is a 1. Otherwise, there is no edge connecting the pair. When generating edges between nodes in different staffing categories, we use similar logic. For example, the probability that there is an edge connecting a nurse with a member of the housekeeping staff is 1.5/16, since, using Table \ref{tab:edges_fitting}, each nurse has an average of 1.5 contacts per day with a member of the housekeeping staff. }
 %That is also the reason that we only need the upper or lower triangular data in \ref{tab:edges_fitting}, we do not want to regenerate edges between two individual. 
 Figure \ref{fig:Example of real-life network} shows an instance of the contact network induced by the procedure just described. %In determining a value of $r_{i,j}$ for any pair of nodes for which an edge exists, we considered the effect of PPE's and social distancing. 
 Our estimates of these $r_{i,j}$ values appear in 
 Table \ref{tab:r_fitting}. For example, if in the contact network there exists an edge between a nurse and a housekeeper, then the probability of transmission (per day) between these two staff members is 0.01. If there is no edge between two staff members, then the transmission probability is 0. We only provide data in the upper part of the table, as this table is symmetric, by our assumptions.

 %On top of that, we also considered the effect of masks and social distances when estimating the probabilities of internal infections that are assigned in different edges. Our final estimation of $r$ are shown in Table \ref{tab:r_fitting}. Note that the contact pattern between two staff $s=i$ and $s=j$ should be reciprocal, which means $r_{i,j}=r_{j,i}$. The upper or lower triangular of data should provide enough information of the heterogeneous probability of internal infection. 
\begin{table}\small
    \centering
    \begin{threeparttable}
        \begin{tabular}{ccccccccccccc}
            \hline
            \textbf{Job title}& \textbf{NR} & \textbf{HK} & \textbf{RH} & \textbf{MS} & \textbf{SL} & \textbf{MT} &\textbf{RC} &\textbf{DN} & \textbf{SC} &\textbf{AD} &\textbf{LG} &\textbf{DR} \\ \hline
            \textbf{NR} &0.08 &0.01 &0.06 &0.015 &0.01 &0.01 &0.04 &0.02 &0.02 &0.01 &0.01 &0.08\\ \hdashline
            \textbf{HK} &$-$  &0.08 &0.01 &0.01 &0.01 &0.01 &0.01 &0.02 &0.02 &0.01 &0.01 &0.04 \\ \hdashline
            \textbf{RH} &$-$ &$-$&0.08 &0.015 &0.01 &0.01 &0.04 &0.02 &0.02 &0.01 &0.01 &0.08\\ \hdashline
            \textbf{MS} &$-$ &$-$ &$-$ &0.08 &0.01 &0.01 &0.04 &0.02 &0.02 &0.01 &0.01 &0.08\\ \hdashline
            \textbf{SL} &$-$ &$-$ &$-$ &$-$ &0.08 &0.01 &0.01 &0.02 &0.02 &0.01 &0.01 &0.08\\ \hdashline
            \textbf{MT} &$-$ &$-$ &$-$ &$-$ &$-$ &0.08 &0.01 &0.02 &0.02 &0.01 &0.01 &0.04\\ \hdashline
            \textbf{RC} &$-$ &$-$ &$-$ &$-$ &$-$ &$-$ &0.08 &0.02 &0.02 &0.01 &0.02 &0.08\\ \hdashline
            \textbf{DN} &$-$ &$-$ &$-$ &$-$ &$-$ &$-$ &$-$ &0.02 &0.04 &0.02 &0.01 &0.08\\ \hdashline
            \textbf{SC} &$-$ &$-$ &$-$ &$-$ &$-$ &$-$ &$-$ &$-$ &0.02 &0.03 &0.02 &0.08\\ \hdashline
            \textbf{AD} &$-$ &$-$ &$-$ &$-$ &$-$ &$-$ &$-$ &$-$ &$-$ &0.08 &0.03 &0.08\\ \hdashline
            \textbf{LG} &$-$ &$-$ &$-$ &$-$ &$-$ &$-$ &$-$ &$-$ &$-$ &$-$ &0.08 &0.08\\ \hdashline
            \textbf{DR} &$-$ &$-$ &$-$ &$-$ &$-$ &$-$ &$-$ &$-$ &$-$ &$-$ &$-$ &0.08\\ \hline
        \end{tabular}
        \begin{tablenotes}
        \item Abbreviations: NR, nurse; HK, house-keeper; MS, mission; SL, salon; RC, receptionist; DN, dining; SC, social; AD, administration; LG, logistic; DR, driver.
        \end{tablenotes}
    \caption{Estimated probability of internal infection among staff in different categories}\label{tab:r_fitting}
\end{threeparttable}
\end{table}

The probability of external infection is a dynamic quantity that depends highly on the current community prevalence of COVID-19. 
In this study, we assume that all staff have the same value of $p$.
%Finally, we fit the parameter for the probability of external infection $p$. At this stage, we assume the staff's social networks are similar outside of the facility. Therefore, we set the same value of $p$ for every staff and that value can be estimated from the most current COVID-19 testing results which is recorded in The New York Times\cite{WB:NYTDatabase}. 
For a given area, such as a state or a county, we can estimate the daily probability of infection for an individual as 
\begin{equation*}
    p=\frac{\text{Number of new infections over the past 7 days}}{7\cdot \text{Population size}}.
\end{equation*}
Of course, the official count new infections in most cases is underestimating the true number of cases. Therefore, in conducting sensitivity analysis, we multiply the estimate above by 
$3$, $5$, and $10$.

% But the recorded newly infected data can be much smaller than the real number, we therefore, multiply $p$ by $3,5$,and $10$ for sensitive analysis.

\subsubsection{Computational experiments}
\begin{figure}[ptbh]
    \centering
    \begin{tikzpicture}
    \begin{axis}[axis lines=left, xlabel=Number of tests per day ($k$), ylabel=Conditional probability of detecting outbreak, width=0.6\textwidth, ymajorgrids=true, xmajorgrids=true, grid style=dotted, legend style={at={(0.98, .02)}, anchor=south east, legend cell align=left}, xtick={0, 10, 20, 30, 40, 50, 60, 70}, xmin=0, ytick={0, 0.2, 0.4, 0.6, 0.8, 1}, ymin=0, ymax=1.01]
    \addplot[color=lightsteelblue,line width=1pt] table[x=k,y=tol3,col sep=comma] {filecontents/K152_c12_p0.003_varyingt.csv};
    \addlegendentry{$t=3$};
    \addplot[color=darkgreen,line width=1pt] table[x=k,y=tol4,col sep=comma] {filecontents/K152_c12_p0.003_varyingt.csv};
    \addlegendentry{$t=4$};
    \addplot[color=chocolate,line width=1pt] table[x=k,y=tol5,col sep=comma] {filecontents/K152_c12_p0.003_varyingt.csv};
    \addlegendentry{$t=5$};
    \addplot[color=basecyan,line width=1pt] table[x=k,y=tol6,col sep=comma] {filecontents/K152_c12_p0.003_varyingt.csv};
    \addlegendentry{$t=6$};
    \addplot[color=baseyellow,line width=1pt] table[x=k,y=tol7,col sep=comma] {filecontents/K152_c12_p0.003_varyingt.csv};
    \addlegendentry{$t=7$};
    \addplot[color=palevioletred,line width=1pt] table[x=k,y=tol8,col sep=comma] {filecontents/K152_c12_p0.003_varyingt.csv};
    \addlegendentry{$t=8$};
    \addplot[color=slateblue,line width=1pt] table[x=k,y=tol9,col sep=comma] {filecontents/K152_c12_p0.003_varyingt.csv};
    \addlegendentry{$t=9$};
    \addplot[color=dimgray,line width=1pt] table[x=k,y=tol10,col sep=comma] {filecontents/K152_c12_p0.003_varyingt.csv};
    \addlegendentry{$t=10$};
    \addplot[color=cssorange,line width=1pt] table[x=k,y=tol11,col sep=comma] {filecontents/K152_c12_p0.003_varyingt.csv};
    \addlegendentry{$t=11$};
    \addplot[color=limegreen,line width=1pt] table[x=k,y=tol12,col sep=comma] {filecontents/K152_c12_p0.003_varyingt.csv};
    \addlegendentry{$t=12$};
    \addplot[color=tan,line width=1pt] table[x=k,y=tol13,col sep=comma] {filecontents/K152_c12_p0.003_varyingt.csv};
    \addlegendentry{$t=13$};
    \addplot[color=crimson,line width=1pt] table[x=k,y=tol14,col sep=comma] {filecontents/K152_c12_p0.003_varyingt.csv};
    \addlegendentry{$t=14$};
    \end{axis}
    \end{tikzpicture}
    \caption{Estimated probabilities of outbreak detection for $K=152$, $c=12$, and $p=0.003$ for various values of $t$}
    \label{fig:real-life various t}
\end{figure}
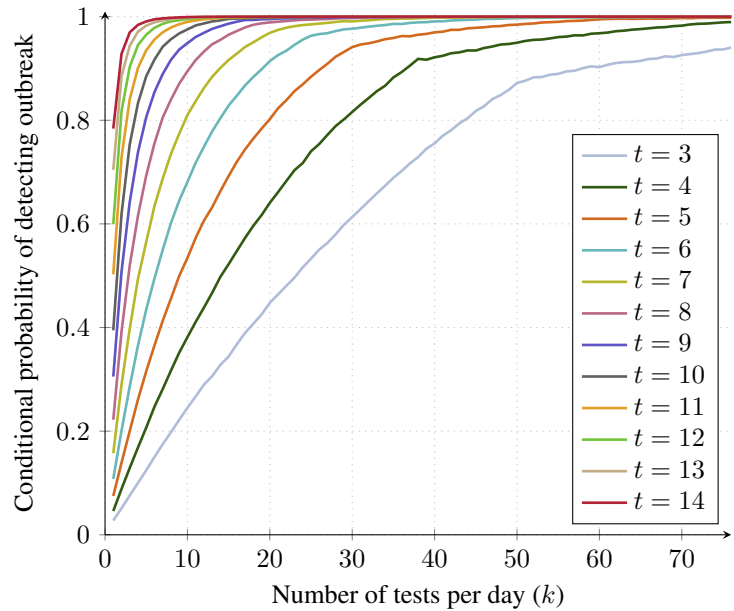

In this section, we present results using the network generated from the data in Tables \ref{tab:raw_data}, \ref{tab:edges_fitting}, and \ref{tab:r_fitting}. Similar to Section \ref{sec:par}, we vary some basic model parameters to perform comparisons. In the first set of experiments, we fix the probability of external infection to $p = 0.003$, and vary the outbreak threshold tolerance $t$.  As shown in Figure \ref{fig:real-life various t}, the results align with the experiments we did for the circulant graph. 
%We can see that as the increase of the outbreak threshold tolerance, the elbow points moves closer to the upper left of the graph.%
Obviously, as before, a larger tolerance value requires a smaller number of tests per day, for a fixed conditional probability target.
For example, if the facility manager targets a 0.8 probability of detecting an outbreak, with $t=3$ then approximately 44 tests per day are required. However, if the threshold of tolerance is set to 4, 5 or even 10 days, then the required number of tests per day is 29, 19 and 3, respectively. We also note for $t=3$ the graph appears to be piecewise linear. This makes intuitive sense, as we do not benefit from the ``network'' effect of virus spread in this case. 
 
%Note that a $t$ days tolerance threshold does not mean the outbreak is detected on the day $t$ after outbreak. Since we test staff everyday, the outbreak can be caught in different days as well. Therefore, it is a good sign for the manager, because it is very likely we can detect the outbreak before our tolerance and the point estimate of average number of days in detecting the outbreak is always smaller than the tolerance $t$.

%\begin{figure} 
%\begin{centering}
%\includegraphics[width=12cm]{Real_case_p_0.003.jpg}
%\caption{Estimated probabilities of outbreak detection for $K=152$, $c=12$, and $p=0.003$ for various values of $t$}  %\label{fig:real-life various t}
%\end{centering}
%\end{figure}

% https://tex.stackexchange.com/questions/215989/controlling-pfplots-axis-labels

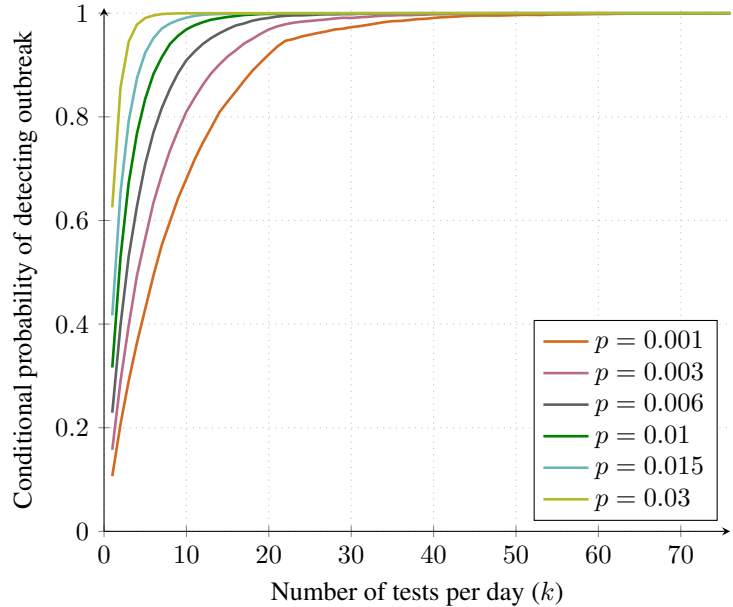
\begin{figure}[ptbh]
    \centering
    \begin{tikzpicture}
    \begin{axis}[axis lines=left, xlabel=Number of tests per day ($k$), ylabel=Conditional probability of detecting outbreak, width=0.6\textwidth, ymajorgrids=true, xmajorgrids=true, grid style=dotted, legend style={at={(0.98, .02)}, anchor=south east, legend cell align=left}, xtick={0, 10, 20, 30, 40, 50, 60, 70}, xmin=0, ytick={0, 0.2, 0.4, 0.6, 0.8, 1}, ymin=0, ymax=1.01]
    \addplot[color=chocolate,line width=1pt] table[x=k,y=prob_p0.001,col sep=comma] {filecontents/K152_c12_t7_varyingp.csv};
    \addlegendentry{$p=0.001$};
    \addplot[color=palevioletred,line width=1pt] table[x=k,y=prob_p0.003,col sep=comma] {filecontents/K152_c12_t7_varyingp.csv};
    \addlegendentry{$p=0.003$};
    \addplot[color=dimgray,line width=1pt] table[x=k,y=prob_p0.006,col sep=comma] {filecontents/K152_c12_t7_varyingp.csv};
    \addlegendentry{$p=0.006$};
    \addplot[color=cssgreen,line width=1pt] table[x=k,y=prob_p0.01,col sep=comma] {filecontents/K152_c12_t7_varyingp.csv};
    \addlegendentry{$p=0.01$};
    \addplot[color=basecyan,line width=1pt] table[x=k,y=prob_p0.015,col sep=comma] {filecontents/K152_c12_t7_varyingp.csv};
    \addlegendentry{$p=0.015$};
    \addplot[color=baseyellow,line width=1pt] table[x=k,y=prob_p0.03,col sep=comma] {filecontents/K152_c12_t7_varyingp.csv};
    \addlegendentry{$p=0.03$};
    \end{axis}
    \end{tikzpicture}
    \caption{Estimated probabilities of outbreak detection in a nursing home case study with $K=152$, $c=12$, $t=7$ and varying values of $p$}
    \label{fig:real-life various p}
\end{figure}

Next, we examine the effect of changing $p$, the 
probability of external infection. 
%As described Section \ref{sec:par}, we estimated $p$ by the data set provided by New York Times, but this estimation can be highly underestimated. For one thing, asymptomatic infections may not been tested. For another, sometimes if one's close family members are infected, the doctors will just claim the one is positive without a test. However, no one knows the exact value for $p$. Therefore, we vary the value of $p$ for sensitivity analysis and run the experiments again. We provide freedom for the stakeholders to choose the value of $p$ that they believe in and provide confidence intervals for the required number of tests. 
Table \ref{fig:real-life various p} shows the results of our simulations. As argued in Section \ref{sec:par}, although it is counterintuitive that higher values of $p$ require fewer tests per day to meet a given probability target, this is actually a logical outcome of the model. 
%For example, if the manager in the nursing home believes the probability of external infection is around $0.006$, and he or she wants to have a 0.9 probability to detect the outbreak within one week. Then, we should look at the yellow line shown in Figure \ref{fig:real-life various p}, it tells around 7 tests per day in this 152 staff nursing home is required, i.e., test all the staff every three weeks, to reach the manager's expectation. 
%\begin{figure} 
%\begin{centering}
%\includegraphics[width=12cm]{Real_case_t0_7.jpg}
%\caption{Estimated probabilities of outbreak detection in a nursing home case study with $K=152$, $c=12$, $t=7$ and varying values of $p$}  \label{fig:real-life various p}
%\end{centering}
%\end{figure}

% https://tex.stackexchange.com/questions/215989/controlling-pfplots-axis-labels
\begin{figure}[ptbh]
    \centering
    \begin{tikzpicture}
    \begin{axis}[axis lines=left, xlabel=Number of days after outbreak, ylabel=Number of staff, width=0.6\textwidth, ymajorgrids=true, xmajorgrids=true, grid style=dotted, legend style={at={(0.02, 0.98)}, anchor=north west, legend cell align=left, font=\footnotesize}, xtick={0, 2, 4, 6, 8, 10, 12, 14, 16, 18, 20}, xmin=0, ytick={0, 20, 40, 60, 80, 100, 120, 140, 160}, ymin=0]
    \addplot[color=royalblue,line width=1pt] table[x=days,y=cuminfected,col sep=comma] {filecontents/K152_c124_p0.003_l3_f0.21.csv};
    \addlegendentry{cumulative infected staff};
    %%% error band %%%
    \addplot[name path=pluserr_infected, draw=none, no markers, forget plot] 
    table[x=days, y=cuminfected95ub, col sep=comma] {filecontents/K152_c124_p0.003_l3_f0.21.csv};
    
    \addplot[name path=minuserr_infected, draw=none, no markers, forget plot] 
    table[x=days, y=cuminfected95lb, col sep=comma] {filecontents/K152_c124_p0.003_l3_f0.21.csv};
    
    \addplot[forget plot, fill=lightskyblue, opacity=0.2]
    fill between[on layer={}, of=pluserr_infected and minuserr_infected];
    %%% error band %%%
    \addplot[color=indianred,line width=1pt] table[x=days,y=cuminfectious,col sep=comma] {filecontents/K152_c124_p0.003_l3_f0.21.csv};
    \addlegendentry{cumulative infectious staff};
    %%% error band %%%
    \addplot[name path=pluserr_infectious, draw=none, no markers, forget plot] 
    table[x=days, y=cuminfectious95ub, col sep=comma] {filecontents/K152_c124_p0.003_l3_f0.21.csv};
    
    \addplot[name path=minuserr_infectious, draw=none, no markers, forget plot] 
    table[x=days, y=cuminfectious95lb, col sep=comma] {filecontents/K152_c124_p0.003_l3_f0.21.csv};
    
    \addplot[forget plot, fill=lightcoral, opacity=0.2]
    fill between[on layer={}, of=pluserr_infectious and minuserr_infectious];
    %%% error band %%%
    \addplot[color=darkgoldenrod,line width=1pt] table[x=days,y=newinfected,col sep=comma] {filecontents/K152_c124_p0.003_l3_f0.21.csv};
    \addlegendentry{newly infected staff};
    %%% error band %%%
    \addplot[name path=pluserr_new, draw=none, no markers, forget plot] 
    table[x=days, y=newinfected95ub, col sep=comma] {filecontents/K152_c124_p0.003_l3_f0.21.csv};
    
    \addplot[name path=minuserr_new, draw=none, no markers, forget plot] 
    table[x=days, y=newinfected95lb, col sep=comma] {filecontents/K152_c124_p0.003_l3_f0.21.csv};
    
    \addplot[forget plot, fill=goldenrod, opacity=0.2]
    fill between[on layer={}, of=pluserr_new and minuserr_new];
    %%% error band %%%
    % ``legend image with text'' defined in the ``preamble''
    \addlegendimage{legend image with text=\ }
    \addlegendentry{confidence interval = 95\%};
    \end{axis}
    \end{tikzpicture}
    \caption{Infection statistics in a case-study model with $K=152$, $c=124$, $p=0.003$, $l=3$, and $f=0.21$}
    \label{fig:infected}
\end{figure}
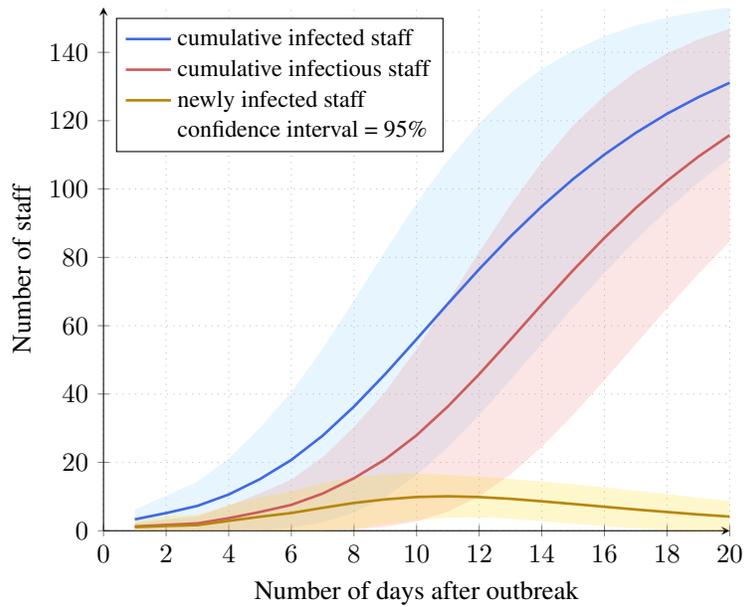

Keeping this observation in mind, we note the a facility manager may want to examine additional metrics when setting the tolerance level $t$ and the detection probability target. For one parameter setting in our case-study network, we provide in Figure \ref{fig:infected} various statistics on the spread on infection, as reflected by 50,000 simulation runs. In Figure \ref{fig:infected}, the blue line represents the cumulative number of infectious staff, the red line represents the number of infected staff, and the yellow line represents the number of newly infected staff. The shaded areas around the lines represent 95\% of confidence intervals around the point estimates for these quantities, various days after the outbreak. As expected, the blue line point estimate is higher than the red line, because infection occurs three days before an individual becomes infectious. Examining the graph, if a facility 
manager wishes to detect an outbreak before 10\% of the staff becomes infected then she would set $t=5$ for the network depicted in Figure \ref{fig:infected}.

%For example, if the manager in the nursing home thinks $p=0.003$ and hopes that around only $1/10$ staff is infected by the time outbreak is detected, he can choose the threshold of tolerance as 5 days. 
%\begin{figure} 
%\begin{centering}
%\includegraphics[width=12cm]{infected_staff.jpg}
%\caption{Infection statistics in a case-study model with $K=152$, $c=124$, $p=0.003$, $l=3$, and $f=0.21$}  %\label{fig:infected}
%\end{centering}
%\end{figure}

% https://tex.stackexchange.com/questions/215989/controlling-pfplots-axis-labels
% https://tex.stackexchange.com/questions/67895/is-there-an-easy-way-of-using-line-thickness-as-error-indicator-in-a-plot

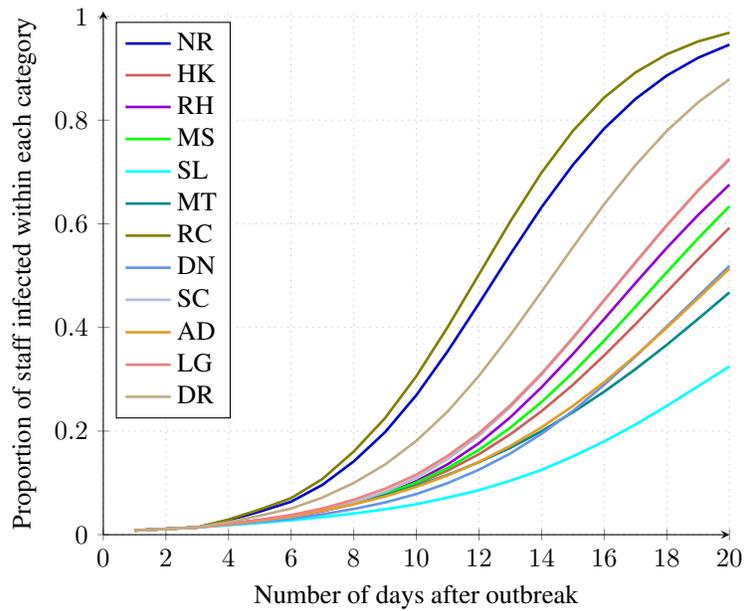
\begin{figure}[ptbh]
    \centering
    \begin{tikzpicture}
    \begin{axis}[axis lines=left, xlabel=Number of days after outbreak, ylabel=Proportion of staff infected within each category, width=0.6\textwidth, ymajorgrids=true, xmajorgrids=true, grid style=dotted, legend style={at={(0.02, 0.98)}, anchor=north west, legend cell align=left}, xtick={0, 2, 4, 6, 8, 10, 12, 14, 16, 18, 20}, xmin=0, ytick={0, 0.2, 0.4, 0.6, 0.8, 1}, ymin=0, ymax=1.01]
    \addplot[color=mediumblue,line width=1pt] table[x=days,y=NR,col sep=comma] {filecontents/s152_c12_p0.003_l3_f0.21.csv};
    \addlegendentry{NR};
    \addplot[color=indianred,line width=1pt] table[x=days,y=HK,col sep=comma] {filecontents/s152_c12_p0.003_l3_f0.21.csv};
    \addlegendentry{HK};
    \addplot[color=darkviolet,line width=1pt] table[x=days,y=RH,col sep=comma] {filecontents/s152_c12_p0.003_l3_f0.21.csv};
    \addlegendentry{RH};
    \addplot[color=csslime,line width=1pt] table[x=days,y=MS,col sep=comma] {filecontents/s152_c12_p0.003_l3_f0.21.csv};
    \addlegendentry{MS};
    \addplot[color=csscyan,line width=1pt] table[x=days,y=SL,col sep=comma] {filecontents/s152_c12_p0.003_l3_f0.21.csv};
    \addlegendentry{SL};
    \addplot[color=darkcyan,line width=1pt] table[x=days,y=MT,col sep=comma] {filecontents/s152_c12_p0.003_l3_f0.21.csv};
    \addlegendentry{MT};
    \addplot[color=cssolive,line width=1pt] table[x=days,y=RC,col sep=comma] {filecontents/s152_c12_p0.003_l3_f0.21.csv};
    \addlegendentry{RC};
    \addplot[color=cornflowerblue,line width=1pt] table[x=days,y=DN,col sep=comma] {filecontents/s152_c12_p0.003_l3_f0.21.csv};
    \addlegendentry{DN};
    \addplot[color=lightsteelblue,line width=1pt] table[x=days,y=SC,col sep=comma] {filecontents/s152_c12_p0.003_l3_f0.21.csv};
    \addlegendentry{SC};
    \addplot[color=cssorange,line width=1pt] table[x=days,y=AD,col sep=comma] {filecontents/s152_c12_p0.003_l3_f0.21.csv};
    \addlegendentry{AD};
    \addplot[color=lightcoral,line width=1pt] table[x=days,y=LG,col sep=comma] {filecontents/s152_c12_p0.003_l3_f0.21.csv};
    \addlegendentry{LG};
    \addplot[color=tan,line width=1pt] table[x=days,y=DR,col sep=comma] {filecontents/s152_c12_p0.003_l3_f0.21.csv};
    \addlegendentry{DR};
    \end{axis}
    \end{tikzpicture}
    \caption{Proportion of infected staff in different job categories with $s=152$, $c=12$, $p=0.003$, $l=3$, and $f=0.21$}
    \label{fig:diff_infection}
\end{figure}

Finally, in Figure \ref{fig:diff_infection} we show the 
proportion of infected staff in various categories over 20 days, again
averaged over 50,000 simulated outbreaks. Note that 
receptionists and nurses are the categories with the highest
infection proportions. In contrast, salon workers have the lowest infection rates in this example. Our discussions with nursing home staff, indicate that their perception is that nurses are quite vulnerable to outbreaks, but receptionists are not. However, if
we scrutinize Table \ref{tab:raw_data} we note that receptionists and nurses both have frequent, close, and prolonged contact with other staff, so the model does seem to be reflecting the data provided.

%To explain this difference, we should look at the data provided in Table \ref{tab:raw_data}. We can tell that receptionists and nurses are the two categories who meet a large amount of other staff, contact duration greater than 15 minutes, and contact distance less than 6 feet. Since we only get access to the self-reported data from one nursing home, the data can be highly objective and bias. Therefore, we believe a more subjective data can lead to more accurate result.

%\begin{figure} 
%\begin{centering}
%\includegraphics[width=12cm]{different_infection.jpg}
%\caption{Proportion of infected staff in different job categories with $s=152$, $c=12$, $p=0.003$, $l=3$, and $f=0.21$}  \label{fig:diff_infection}
%\end{centering}
%\end{figure}

% https://tex.stackexchange.com/questions/215989/controlling-pfplots-axis-labels
\subsection{Effect of Testing Order}
In Section \ref{sec:testo}, we examined the effects of testing order in what we called homogeneous graphs, i.e., graphs with a significant amount symmetry and homogeneity among the nodes. There, we found small, but mostly insignificant differences among testing protocols. In contact networks arising from the real world, the associated graphs are not expected to have such a regular structure. Due to the less regular structure, there are also a larger variety of heuristic testing protocols that can be tested. The most intuitive protocols are based on testing ``central'' or important nodes first. 

In this section, we test three such heuristics of varying levels of complexity. In each heuristic we need to produce an ordered list (permutation) of the nodes that determines the testing protocol. The first heuristic, which we call \emph{degree rank}, we simply order the nodes from highest degree to lowest degree, breaking ties in an arbitrary manner. Nodes are then tested in this order. The idea, of course, is a higher degree node is more likely to ``detect'' an outbreak since it has connections to lots of other nodes. The second heuristic, which is called \emph{PageRank}, is based on the well-known node ranking system first developed by the founders of Google. We compute the PageRank for all nodes, and test in the order of highest rank to lowest. Again, the idea of this ranking is that higher ranked nodes would be visited more often by a virus surfing the contact network. A potential drawback of this notion, as applied to surveillance testing, is that PageRank is essentially derived from the steady-state distribution of the virus surfing process alluded to above. However, in our virus spread and detection model, we are really concerned with the \emph{transient} behavior of the virus. In particular, what is most important for early detection is the initial trajectory of the virus over the network. This observation inspires us to consider another algorithm which we call 
\emph{simulation rank}. The idea is as follows: if our outbreak tolerance is $t$ days, then we are interested in testing the nodes that are most likely to be involved in an outbreak during the $t$ days. In particular, we should rank the nodes according to the probability that they are part of a $t$-day outbreak. In theory, one could compute this probability exactly, but it is clearly an intractable computation for networks of even moderate size. Instead, we simulate many virus sample paths to estimate the aforementioned probability for each node, and then use these estimates to create the ranking. It is clear that the degree rank ordering requires very little computational effort, as it just involves counting edges. PageRank is more difficult, but there are efficient algorithms to estimate the PageRank, even for large networks. Finally, simulation rank requires a considerable amount of effort, as potentially thousands of simulation samples are required to get a good estimate. 

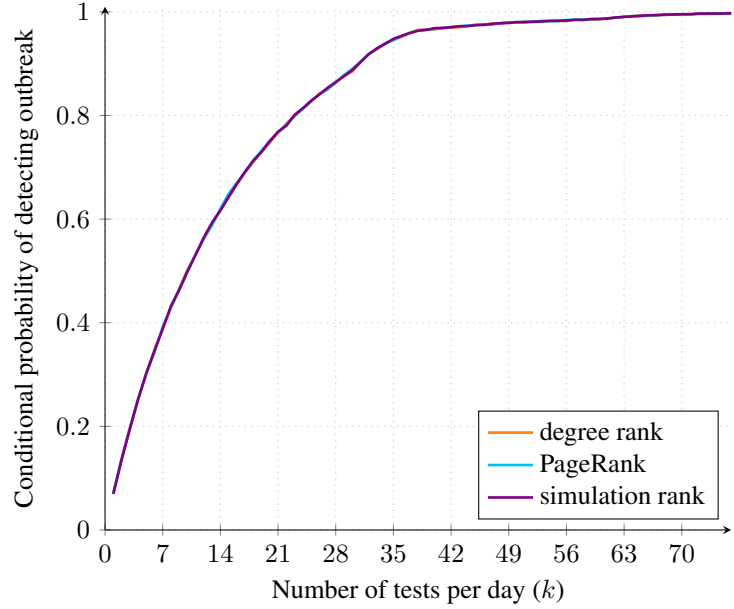
\begin{figure}[ptbh]
    \centering
    \begin{tikzpicture}
        \begin{axis}[axis lines=left, xlabel=Number of tests per day ($k$), ylabel=Conditional probability of detecting outbreak, width=0.6\textwidth, ymajorgrids=true, xmajorgrids=true, grid style=dotted, legend style={at={(0.98, .02)}, anchor=south east, legend cell align=left}, xtick={0, 7, 14, 21, 28, 35, 42, 49, 56, 63, 70}, xmin=0, ytick={0, 0.2, 0.4, 0.6, 0.8, 1}, ymin=0, ymax=1.01]
        \addplot[color=taborange,line width=1pt] table[x=k, y=degreerank_prob, col sep=comma] {filecontents/LTC_s152_c12_p0.006_l3_t4_f0.21_varyingtestprotocol.csv};
        \addlegendentry{degree rank};
        \addplot[color=deepskyblue,line width=1pt] table[x=k, y=PageRank_prob, col sep=comma] {filecontents/LTC_s152_c12_p0.006_l3_t4_f0.21_varyingtestprotocol.csv};
        \addlegendentry{PageRank};
        \addplot[color=csspurple,line width=1pt] table[x=k, y=simulationrank_prob, col sep=comma] {filecontents/LTC_s152_c12_p0.006_l3_t4_f0.21_varyingtestprotocol.csv};
        \addlegendentry{simulation rank};
        \end{axis}
    \end{tikzpicture}
    \caption{Performance of heuristic testing protocols for the LTC network with $s=152$, $c=12$, $p=0.006$, $l=3$, $t=4$ and $f=0.21$}
    \label{fig:real4}
\end{figure}

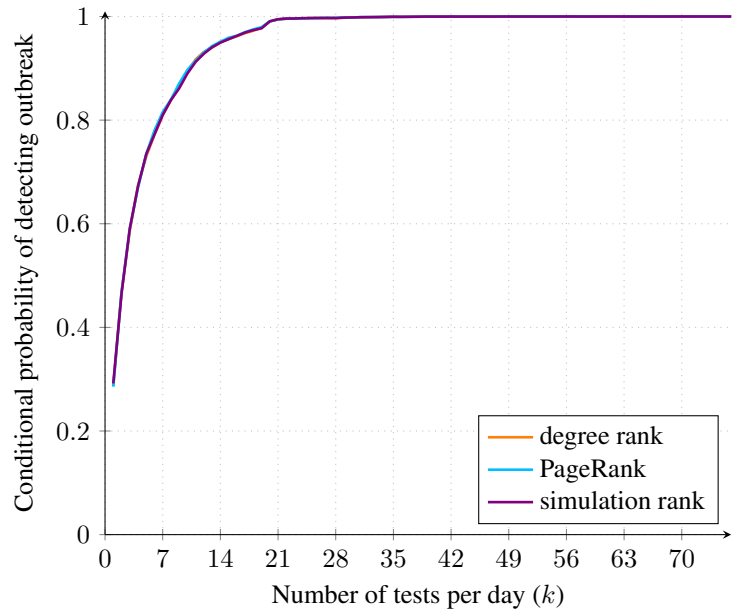
\begin{figure}[ptbh]
    \centering
    \begin{tikzpicture}
        \begin{axis}[axis lines=left, xlabel=Number of tests per day ($k$), ylabel=Conditional probability of detecting outbreak, width=0.6\textwidth, ymajorgrids=true, xmajorgrids=true, grid style=dotted, legend style={at={(0.98, .02)}, anchor=south east, legend cell align=left}, xtick={0, 7, 14, 21, 28, 35, 42, 49, 56, 63, 70}, xmin=0, ytick={0, 0.2, 0.4, 0.6, 0.8, 1}, ymin=0, ymax=1.01]
        \addplot[color=taborange,line width=1pt] table[x=k, y=degreerank_prob, col sep=comma] {filecontents/LTC_s152_c12_p0.006_l3_t7_f0.21_varyingtestprotocol.csv};
        \addlegendentry{degree rank};
        \addplot[color=deepskyblue,line width=1pt] table[x=k, y=PageRank_prob, col sep=comma] {filecontents/LTC_s152_c12_p0.006_l3_t7_f0.21_varyingtestprotocol.csv};
        \addlegendentry{PageRank};
        \addplot[color=csspurple,line width=1pt] table[x=k, y=simulationrank_prob, col sep=comma] {filecontents/LTC_s152_c12_p0.006_l3_t7_f0.21_varyingtestprotocol.csv};
        \addlegendentry{simulation rank};
        \end{axis}
    \end{tikzpicture}
    \caption{Performance of heuristic testing protocols for the LTC network with $s=152$, $c=12$, $p=0.006$, $l=3$, $t=7$ and $f=0.21$}
    \label{fig:real7}
\end{figure}

In Figures \ref{fig:real4} and \ref{fig:real7}, we show the results of testing the three ranking algorithms for test protocols. In each case, we determine the test order and then evaluate the protocol using the network depicted in Figure \ref{fig:Example of real-life network}. As in our previous computations, we using 50,000 virus sample paths to perform the evaluation. Figure \ref{fig:real4} displays the results for a tolerance of $t=4$ days. Clearly, the difference among the algorithms is imperceptible in this figure. We also ran statistical tests confirming that there no statistically significant difference among the algorithms. Figure 
\ref{fig:real7} displays the results for a tolerance of $t=7$ days. Again, the differences are almost negligible, although PageRank appears to slightly outperform the other algorithms when 7 to 10 tests per days are performed. Still, these differences do not appear to be statistically significant. We performed similar tests on a small set of other networks, with similar results. Although the tests are not comprehensive, they seem to indicate that a relatively simple algorithm such as degree rank works just as well as more sophisticated algorithms. 

%\begin{figure} 
%\begin{centering}
%\includegraphics[width=12cm]{real_t=4.png}
%\caption{Performance of heuristic testing protocols for the LTC network with $s=152$, $c=12$, $p=0.006$, $l=3$, $t=4$ and $f=0.21$}  \label{fig:real4}
%\end{centering}
%\end{figure}

%% https://tex.stackexchange.com/questions/215989/controlling-pfplots-axis-labels

%\begin{figure} 
%\begin{centering}
%\includegraphics[width=12cm]{real_t=7.png}
%\caption{Performance of heuristic testing protocols for the LTC network with $s=152$, $c=12$, $p=0.006$, $l=3$, $t=7$ and $f=0.21$}  \label{fig:real7}
%\end{centering}
%\end{figure}

%% https://tex.stackexchange.com/questions/215989/controlling-pfplots-axis-labels

\pagebreak
\bibliographystyle{abbrv}
%\bibliography{has}

\begin{thebibliography}{10}

\bibitem{Albogami2021}
Y.~Albogami, H.~Alkofide, and A.~Alrwisan.
\newblock {Covid-19 Vaccine Surveillance in Saudi Arabia: Opportunities for
  Real-time Assessment}.
\newblock {\em Saudi Pharmaceutical Journal}, 29(8):914--916, 2021.

\bibitem{Ames2011}
G.~M. Ames, D.~B. George, C.~P. Hampson, A.~R. Kanarek, C.~D. Mcbee, D.~R.
  Lockwood, J.~D. Achter, and C.~T. Webb.
\newblock {Using network properties to predict disease dynamics on human
  contact networks}.
\newblock {\em Proceedings of the Royal Society B: Biological Sciences},
  278(1724):3544--3550, 2011.

\bibitem{Baek2020}
Y.~J. Baek, T.~Lee, Y.~Cho, J.~H. Hyun, M.~H. Kim, Y.~Sohn, J.~H. Kim, J.~Y.
  Ahn, S.~J. Jeong, N.~S. Ku, J.~S. Yeom, J.~Lee, and J.~Y. Choi.
\newblock {A mathematical model of COVID-19 transmission in a tertiary hospital
  and assessment of the effects of different intervention strategies}.
\newblock {\em PLoS ONE}, 15(10):1--16, 2020.

\bibitem{Bai2017}
Y.~Bai, B.~Yang, L.~Lin, J.~L. Herrera, Z.~Du, and P.~Holme.
\newblock {Optimizing sentinel surveillance in temporal network epidemiology}.
\newblock {\em Scientific Reports}, 7(1):1--10, 2017.

\bibitem{Bernadou2021}
A.~Bernadou, S.~Bouges, M.~Catroux, J.~C. Rigaux, C.~Laland, N.~Lev{\^{e}}que,
  U.~Noury, S.~Larrieu, S.~Acef, D.~Habold, F.~Cazenave-Roblot, and L.~Filleul.
\newblock {High impact of COVID-19 outbreak in a nursing home in the
  Nouvelle-Aquitaine region, France, March to April 2020}.
\newblock {\em BMC Infectious Diseases}, 21(1):1--6, 2021.

\bibitem{Besse2021}
C.~Besse and G.~Faye.
\newblock {Dynamics of epidemic spreading on connected graphs}.
\newblock {\em Journal of Mathematical Biology}, 82(6):1--52, 2021.

\bibitem{Chapman2021}
L.~A. Chapman, M.~Kushel, S.~N. Cox, A.~Scarborough, C.~Cawley, T.~Q. Nguyen,
  I.~Rodriguez-Barraquer, B.~Greenhouse, E.~Imbert, and N.~C. Lo.
\newblock {Comparison of infection control strategies to reduce COVID-19
  outbreaks in homeless shelters in the United States: a simulation study}.
\newblock {\em BMC Medicine}, 19(1):1--13, 2021.

\bibitem{ChenHJ2021}
H.~J. Chen, H.~J. Lin, M.~C. Wu, H.~J. Tang, B.~A. Su, and C.~C. Lai.
\newblock {The implementation of an active surveillance integrating information
  technology and drive-through coronavirus testing station for suspected
  COVID-19 cases}.
\newblock {\em Journal of Infection}, 82(2):282--327, 2021.

\bibitem{Chen2020}
L.~Chen and F.~Wei.
\newblock {Study on a susceptible–exposed–infected–recovered model with
  nonlinear incidence rate}.
\newblock {\em Advances in Difference Equations}, 2020(206), 2020.

\bibitem{Chowell2016}
G.~Chowell, L.~Sattenspiel, S.~Bansal, and C.~Viboud.
\newblock {Mathematical models to characterize early epidemic growth: A
  review}.
\newblock {\em Physics of Life Reviews}, 18:66--97, 2016.

\bibitem{Ciaperoni2020}
M.~Ciaperoni, E.~Galimberti, F.~Bonchi, C.~Cattuto, F.~Gullo, and A.~Barrat.
\newblock {Relevance of temporal cores for epidemic spread in temporal
  networks}.
\newblock {\em Scientific Reports}, 10(1):1--15, 2020.

\bibitem{Drenkard2021}
K.~Drenkard, B.~Sakallaris, P.~Deyo, S.~Abdillahi, and H.~Hahn.
\newblock {University COVID-19 Surveillance Testing Center: Challenges and
  Opportunities for Schools of Nursing}.
\newblock {\em Journal of Professional Nursing}, 37(5):948--953, 2021.

\bibitem{duval2018measuring}
A.~Duval, T.~Obadia, L.~Martinet, P.-Y. Bo{\"e}lle, E.~Fleury, D.~Guillemot,
  L.~Opatowski, and L.~Temime.
\newblock Measuring dynamic social contacts in a rehabilitation hospital:
  {E}ffect of wards, patient and staff characteristics.
\newblock {\em Scientific reports}, 8(1):1--11, 2018.

\bibitem{Ferretti2020}
L.~Ferretti, C.~Wymant, M.~Kendall, L.~Zhao, A.~Nurtay, L.~Abeler-D{\"{o}}rner,
  M.~Parker, D.~Bonsall, and C.~Fraser.
\newblock {Quantifying SARS-CoV-2 transmission suggests epidemic control with
  digital contact tracing}.
\newblock {\em Science}, 368(6491), 2020.

\bibitem{Gaeta2021}
G.~Gaeta.
\newblock {A simple SIR model with a large set of asymptomatic infectives}.
\newblock {\em Mathematics In Engineering}, 3(2):1--39, 2021.

\bibitem{Garibaldi2021}
P.~M. Garibaldi, N.~N. Ferreira, G.~R. Moraes, J.~C. Moura, D.~L.
  Esp{\'{o}}sito, G.~J. Volpe, R.~T. Calado, B.~A. Fonseca, and M.~C. Borges.
\newblock {Efficacy of COVID-19 outbreak management in a skilled nursing
  facility based on serial testing for early detection and control}.
\newblock {\em Brazilian Journal of Infectious Diseases}, 25(2):1--6, 2021.

\bibitem{Harada2020}
S.~Harada, S.~Uno, T.~Ando, M.~Iida, Y.~Takano, Y.~Ishibashi, Y.~Uwamino,
  T.~Nishimura, A.~Takeda, S.~Uchida, A.~Hirata, M.~Sata, M.~Matsumoto,
  A.~Takeuchi, H.~Obara, H.~Yokoyama, K.~Fukunaga, M.~Amagai, Y.~Kitagawa,
  T.~Takebayashi, and N.~Hasegawa.
\newblock {Control of a Nosocomial Outbreak of COVID-19 in a University
  Hospital}.
\newblock {\em Open Forum Infectious Diseases}, 7(12):1--9, 2020.

\bibitem{Herrera2016}
J.~L. Herrera, R.~Srinivasan, J.~S. Brownstein, A.~P. Galvani, and L.~A.
  Meyers.
\newblock {Disease Surveillance on Complex Social Networks}.
\newblock {\em PLoS Computational Biology}, 12(7):1--16, 2016.

\bibitem{herrera2021network}
J.~L. Herrera-Diestra, M.~Tildesley, K.~Shea, and M.~Ferrari.
\newblock Network structure and disease risk for an endemic infectious disease.
\newblock {\em arXiv preprint arXiv:2107.06186}, 2021.

\bibitem{Hou2020}
C.~Hou, J.~Chen, Y.~Zhou, L.~Hua, J.~Yuan, S.~He, Y.~Guo, S.~Zhang, Q.~Jia,
  C.~Zhao, J.~Zhang, G.~Xu, and E.~Jia.
\newblock {The effectiveness of quarantine of Wuhan city against the Corona
  Virus Disease 2019 (COVID-19): A well-mixed SEIR model analysis}.
\newblock {\em Journal of Medical Virology}, 92(7):841--848, 2020.

\bibitem{Leal-Neto2020}
O.~B. Leal-Neto, F.~A. Santos, J.~Y. Lee, J.~O. Albuquerque, and W.~V. Souza.
\newblock {Prioritizing COVID-19 tests based on participatory surveillance and
  spatial scanning}.
\newblock {\em International Journal of Medical Informatics}, 143:104263, 2020.

\bibitem{mastin2020optimising}
A.~J. Mastin, T.~R. Gottwald, F.~van~den Bosch, N.~J. Cunniffe, and S.~Parnell.
\newblock Optimising risk-based surveillance for early detection of invasive
  plant pathogens.
\newblock {\em PLoS biology}, 18(10):e3000863, 2020.

\bibitem{McMichael2020}
T.~M. McMichael, D.~W. Currie, S.~Clark, S.~Pogosjans, M.~Kay, N.~G. Schwartz,
  J.~Lewis, A.~Baer, V.~Kawakami, M.~D. Lukoff, J.~Ferro, C.~Brostrom-Smith,
  T.~D. Rea, M.~R. Sayre, F.~X. Riedo, D.~Russell, B.~Hiatt, P.~Montgomery,
  A.~K. Rao, E.~J. Chow, F.~Tobolowsky, M.~J. Hughes, A.~C. Bardossy, L.~P.
  Oakley, J.~R. Jacobs, N.~D. Stone, S.~C. Reddy, J.~A. Jernigan, M.~A. Honein,
  T.~A. Clark, and J.~S. Duchin.
\newblock {Epidemiology of Covid-19 in a Long-Term Care Facility in King
  County, Washington}.
\newblock {\em New England Journal of Medicine}, 382(21):2005--2011, 2020.

\bibitem{Meyers2005}
L.~A. Meyers, B.~Pourbohloul, M.~E. Newman, D.~M. Skowronski, and R.~C.
  Brunham.
\newblock {Network theory and SARS: Predicting outbreak diversity}.
\newblock {\em Journal of Theoretical Biology}, 232(1):71--81, 2005.

\bibitem{Rennert2021}
L.~Rennert, C.~McMahan, C.~A. Kalbaugh, Y.~Yang, B.~Lumsden, D.~Dean,
  L.~Pekarek, and C.~C. Colenda.
\newblock {Surveillance-based informative testing for detection and containment
  of SARS-CoV-2 outbreaks on a public university campus: an observational and
  modelling study}.
\newblock {\em The Lancet Child and Adolescent Health}, 5(6):428--436, 2021.

\bibitem{Saha2021}
S.~Saha and S.~Saha.
\newblock {The impact of the undetected COVID-19 cases on its transmission
  dynamics}.
\newblock {\em Indian Journal of Pure and Applied Mathematics}, pages 1--6,
  June 2021.

\bibitem{Shaikh2020}
A.~S. Shaikh, I.~N. Shaikh, and K.~S. Nisar.
\newblock {A mathematical model of COVID-19 using fractional derivative:
  outbreak in India with dynamics of transmission and control}.
\newblock {\em Advances in Difference Equations}, 2020(1), 2020.

\bibitem{Shu2012}
H.~Shu, D.~Fan, and J.~Wei.
\newblock {Global stability of multi-group SEIR epidemic models with
  distributed delays and nonlinear transmission}.
\newblock {\em Nonlinear Analysis: Real World Applications}, 13(4):1581--1592,
  2012.

\bibitem{Smith2020best}
D.~R. Smith, A.~Duval, K.~B. Pouwels, D.~Guillemot, J.~Fernandes, B.-T. Huynh,
  L.~Temime, and L.~Opatowski.
\newblock How best to use limited tests? {I}mproving {COVID}-19 surveillance in
  long-term care.
\newblock {\em medRxiv}, 2020.

\bibitem{Smith2020}
D.~R. Smith, A.~Duval, K.~B. Pouwels, D.~Guillemot, J.~Fernandes, B.~T. Huynh,
  L.~Temime, and L.~Opatowski.
\newblock {Optimizing COVID-19 surveillance in long-term care facilities: a
  modelling study}.
\newblock {\em BMC Medicine}, 18(1):1--16, 2020.

\bibitem{Strand2021}
R.~Strand, N.~Fernstr{\"{o}}m, A.~Holmberg, Y.~{De Marinis}, C.~J. Fraenkel,
  and M.~Rasmussen.
\newblock {Post-outbreak serological screening for SARS-CoV-2 infection in
  healthcare workers at a Swedish University Hospital}.
\newblock {\em Infectious Diseases}, 53(9):707--712, 2021.

\bibitem{Sudre2021}
C.~H. Sudre, A.~Keshet, M.~S. Graham, A.~D. Joshi, S.~Shilo, H.~Rossman,
  B.~Murray, E.~Molten, K.~Klaser, L.~D. Canas, M.~Antonelli, L.~H. Nguyen,
  D.~A. Drew, M.~Modat, J.~C. Pujol, S.~Ganesh, J.~Wolf, T.~Meir, A.~T. Chan,
  C.~J. Steves, T.~D. Spector, J.~S. Brownstein, E.~Segal, S.~Ourselin, and
  C.~M. Astley.
\newblock {Anosmia, ageusia, and other COVID-19-like symptoms in association
  with a positive SARS-CoV-2 test, across six national digital surveillance
  platforms: an observational study.}
\newblock {\em The Lancet. Digital health}, 7500(21):1--10, 2021.

\bibitem{danis2020high}
E.~P. H.~E. Team, K.~Danis, L.~Fonteneau, S.~Georges, C.~Daniau,
  S.~Bernard-Stoecklin, L.~Domegan, J.~O’Donnell, S.~H. Hauge, S.~Dequeker,
  E.~Vandael, J.~Van~der Heyden, F.~Renard, N.~B. Sierra, E.~Ricchizzi,
  B.~Schweickert, N.~Schmidt, M.~Abu~Sin, T.~Eckmanns, J.~Paiva, and
  E.~Schneider.
\newblock {High impact of COVID-19 in long-term care facilities, suggestion for
  monitoring in the EU/EEA, May 2020}.
\newblock {\em Eurosurveillance}, 25(22):2000956, 2020.

\bibitem{10.1093/cid/ciaa682}
L.~Temime, M.-P. Gustin, A.~Duval, N.~Buetti, P.~Crépey, D.~Guillemot,
  R.~Thiébaut, P.~Vanhems, J.-R. Zahar, D.~R.~M. Smith, and L.~Opatowski.
\newblock A conceptual discussion about the basic reproduction number of severe
  acute respiratory syndrome coronavirus 2 in healthcare settings.
\newblock {\em Clinical Infectious Diseases}, 72(1):141--143, January 2021.

\bibitem{Yen2020}
M.~Y. Yen, J.~Schwartz, C.~C. King, C.~M. Lee, and P.~R. Hsueh.
\newblock {Recommendations for protecting against and mitigating the COVID-19
  pandemic in long-term care facilities}.
\newblock {\em Journal of Microbiology, Immunology and Infection},
  53(3):447--453, 2020.

\end{thebibliography}

\end{document}